\documentstyle[12pt]{article}
\def\figa#1#2#3{
  \begin{figure}[htb]
  \epsfxsize=#2in
  \centerline{\epsffile{#1.ps}}
  \caption{#3}
  \end{figure}
}
\def\figp#1#2#3{
  \begin{figure}[htbp]
  \epsfxsize=#2in
  \centerline{\epsffile{#1.ps}}
  \caption{#3}
  \end{figure}
}
\input epsf
\newcommand{\lesssim}{
 \mathrel{\setbox0=\hbox{$<$}\raise0.6ex\copy0\kern-\wd0
 \lower0.65ex\hbox{$\sim$}}}
\newcommand{\gtrsim}{
 \mathrel{\setbox0=\hbox{$>$}\raise0.6ex\copy0\kern-\wd0
 \lower0.65ex\hbox{$\sim$}}}

\begin{document}

\title{DIFFRACTIVE HEAVY QUARKONIUM PHOTO- AND ELECTROPRODUCTION IN QCD}

\author{ Leonid Frankfurt\thanks{\noindent On leave of absence from the 
         St.Petersburg Nuclear Physics Institute, Russia.}\\
         School of Physics and Astronomy\\ 
         Raymond and Beverly Sackler Faculty of 
         Exact Sciences\\
         Tel Aviv University, Tel Aviv, Israel
\\[0.3cm]
         Werner Koepf\thanks{\noindent Now at NeuralWare, a subsidiary of Aspen
         Technology Inc.}\\
         Department of Physics, The Ohio State University\\
         Columbus, OH 43210, USA
\\[0.3cm]
         Mark Strikman\thanks{\noindent Also at St.Petersburg Nuclear Physics 
          Institute, Russia.}\\
         Department of Physics, Pennsylvania State University\\
         University Park, PA 16802, USA
\\[0.6cm]}

\date{(revised September 10, 1997)}

\maketitle

\vspace{-18.5cm}
\begin{flushleft}
\tt DESY-97-028, OSU-97-0201 \\ February 1997
\end{flushleft}
\vspace{15.5cm}

\begin{abstract}
Hard diffractive photo- and electroproduction of heavy vector mesons 
($J/\psi$ and $\Upsilon$) is evaluated within the leading $\alpha_s\ln{Q^2
\over\Lambda_{QCD}^2}$ approximation of QCD.  In difference from our
earlier work on that subject, also the production of transversely
polarized vector mesons is calculated.  Special emphasis is placed on 
the role of the vector meson's $q\bar q$ light-cone wave function.  
In that context, conventional non-relativistic quarkonium models and
a light-front QCD bound state calculation are critically examined and 
confronted with QCD expectations.  Our numerical analysis finds a 
significant high momentum tail in the latter wave functions and a 
deviation from the expected asymptotic behavior of $\phi_V(z,b=0)\propto 
z(1-z)$.  We then design an interpolation to match the quarkonium models 
at large inter-quark separations with QCD expectations at small distances.  
We use these results to compare our predictions for the forward 
differential cross section of $J/\psi$ photo- and electroproduction with 
recent experimental results from HERA.  In addition, our earlier discussion 
of $\rho^o$ electroproduction is updated in light of recent experimental 
and theoretical enhancements.
\end{abstract}

\section{Introduction}
\label{I}

Diffractive vector meson production opens a precious window on the interface
between perturbative QCD and hadronic physics.  While elastic processes are
commonly described through non-perturbative, phenomenological methods, as, for 
instance, soft Pomeron exchange \cite{pom}, hard inclusive reactions -- most 
prominently deep inelastic lepton scattering -- are, in a sense, exactly 
calculable 
as a consequence of the QCD factorization theorem.
These two classes 
of processes now meet at HERA.  However, similar to inclusive deep inelastic 
scattering, also the amplitude for diffractive (coherent) production of vector 
mesons in deep inelastic lepton-nucleon scattering factorizes into a hard part 
calculable in pQCD convoluted with the non-perturbative off-diagonal gluon 
distribution in the target \cite{Brod94}.  A rigorous QCD-based proof of the
factorization theorem for hard exclusive electroproduction of vector mesons,
valid to all orders in perturbation theory, was recently given in Ref. 
\cite{Collins}.  This theorem holds if only short distances contribute, 
which is the case for the production of longitudinally polarized $\rho^o$ 
at sufficiently large $Q^2$ or heavy flavor photo- and electroproduction 
\cite{hepph}.

For large but non-asymptotic 
photon virtuality, 
 the hard amplitude for
exclusive vector meson production is sensitive to the transverse momentum 
distribution in the light-cone wave function of the $q\bar q$ leading Fock 
component of the produced vector meson \cite{hepph}.  This leads to a 
suppression of the asymptotic amplitude, i.e., to an interplay between the 
quark(antiquark) momentum distribution in the vector meson and the $Q^2$ 
dependence of the corresponding cross section.  That, in turn, allows to 
extract information on this wave function -- and hence on the three dimensional
distribution of color in the produced hadron -- from the $Q^2$ 
and the $t$ dependences
of the cross section.

In this work, we focus the QCD analysis of Refs. \cite{Brod94} and \cite{hepph}
on heavy quarkonium ($J/\psi$ and $\Upsilon$) photo- and electroproduction.  
Furthermore, we extend the respective formalism, which in Refs. \cite{Brod94}
and \cite{hepph} was applied to the production of longitudinally polarized 
vector mesons only, to transverse polarizations as well.  The important role 
the vector meson's $q\bar q$ light-cone wave function plays in diffractive
photo- and electroproduction at non-asymptotic $Q^2$ requires a detailed study 
of this quantity.  Motivated by the large value of the quark mass in heavy 
quarkonia, we start from conventional non-relativistic potential models 
\cite{charmlog,charmqcd,charmpow,charmcor} and/or a non-relativistic 
light-front QCD bound state calculation \cite{perry}.  We then critically 
examine the respective wave functions and confront them with QCD 
expectations.  

In particular for the $J/\psi$ meson, our numerical analysis yields a 
significant value for the high momentum component in the respective 
non-relativistic wave functions $\phi_V(k)$.  For instance for the potential 
model of Ref. \cite{charmlog}, the region ${v \over c} \geq 1$, where the 
non-relativistic approximation is definitely inadequate, contributes
over 30\% to the integral $\int\!d^3k\,\phi_V(k)$.  Latter integral appears in the 
expression for the $V \rightarrow e^+e^-$ decay width.  This is illustrated 
in Fig. 6, and it is in line with the QCD prediction of large relativistic
corrections to the corresponding bound state equations \cite{buchmueller}.
Those large relativistic effects put the validity of a non-relativistic
description of $J/\psi$ mesons -- and, in particular, a non-relativistic
evaluation 
of their production in high energy processes -- seriously into question. 
Our analysis shows that the $Q^2$ dependence of $J/\psi$ electroproduction, 
the photoproduction cross section ratios of $\Upsilon$ and $J/\psi$ mesons,
and modifications in the $t$ slope of those cross sections
are good probes for the color distribution in the light-cone wave function of 
the vector mesons as well as the dependence of the parton 
distribution in the target on the produced meson's transverse size.
In particular, these effects lead to an enhancement of the 
cross section ratio for diffractive electroproduction of $\Upsilon$ and 
$J/\psi$ mesons by a factor $\approx 10$ for the same $x$
as compared to the naive scaling estimate.  This was
discussed already in Ref. \cite{hepph}.

In addition, if we express the non-relativistic wave functions in
terms of light-cone coordinates, we find that they do not display the 
expected asymptotic behavior \cite{CZ} $\int\!d^2k_t\,
\phi_{V~L}(z,k_t) \propto z(1-z)$
in the vicinity of $z=0$ or $z=1$.  
This is illustrated in Fig. 8.  Another mismatch between the 
non-relativistic and the light-cone approach appears within the evaluation of 
the $V \rightarrow e^+e^-$ decay width.  When $\Gamma_{V \rightarrow e^+e^-}$ 
is calculated from the non-relativistic wave function $\phi_V(k)$, 
a QCD correction factor, $1-{16\alpha_s\over 3\pi}$, appears \cite{barbieri},
which can be numerically large (${16\alpha_s\over 3\pi}\approx 0.5$ 
for $J/\psi$ where we use $\alpha_S(J/\psi)=0.3$)
while no such term is present  in the relation
\cite{LastXX} with the light-cone $q\bar q$ wave function $\phi_V(z,k_t)$.
This difference may be important in practice since the Schwinger formula 
for the positronium decay \cite{Schwinger} becomes inaccurate 
for charmonium where the high momentum component in the 
wave function is not small. 
To remedy these deficiencies, we designed an interpolation for the wave 
function of heavy quarkonia which smoothly matches the wave functions 
obtained at average inter-quark separations from non-relativistic potential 
models (or within a light-front QCD bound state calculation) with QCD 
predictions at small distances.  

The basic difference of the current work from Ref. \cite{hepph} is 
that the formulae
valid in leading order in $1 \over Q^2 + 4m^2$ are derived
by decomposing Feynman diagrams over the {\em transverse} distance between
bare quarks, and that the quarkonium light-cone wave functions which we used
respect QCD 
predictions for their high momentum tail.  As for any hard process, the cross 
section is expressed through the distribution of bare quarks in the vector 
meson and not through the distribution of constituent quarks, as 
it has been assumed 
in Refs. \cite{Ryskin} and \cite{Ryskin2}.  In the latter investigations,
the cross section for diffractive photo- and electroproduction of 
$J/\psi$ mesons was evaluated
in the BFKL approximation and while employing a non-relativistic 
constituent quark model.  No corrections arising from the quark motion 
within the produced $J/\psi$ mesons were considered in Ref. \cite{Ryskin}.  
In a later work \cite{Ryskin2}, the authors then argued that the respective 
corrections are small within realistic charmonium models.  This is at 
variance with our findings.  In addition, our numerical analysis shows that 
the static approximation used in Refs. \cite{Ryskin} and \cite{Ryskin2} 
is not in line with conventional charmonium models. 
Neglect of quark Fermi motion and related color screening
effects in Refs. \cite{Ryskin,Ryskin2} leads to factor $\approx 3$ 
suppression for the ratio of cross sections of photoproduction of 
$\Upsilon$ versus $J/\psi$ mesons as compared to the results in this paper.

After outlining the basic formalism in Sect. \ref{II}, in Sect. \ref{III}, 
we discuss the heavy vector meson's light-cone wave function which 
describes its leading $q\bar q$ Fock state component.  We then 
compare, in Sect. \ref{IV}, with recent 
experimental results from HERA for $J/\psi$ photo- \cite{H1phot} and 
electroproduction \cite{H1}.  In Sect. \ref{V}, we update our discussion of
$\rho^o$ electroproduction in light of recent experimental \cite{ZEUS94} and
theoretical \cite{halperin} enhancements.  We summarize and conclude in Sect. 
\ref{VI}.

\section{The basic formalism}
\label{II}

\subsection{The forward differential cross section}
\label{II.A}

In Ref. \cite{Brod94}, the forward differential cross section for the 
production of longitudinally polarized vector mesons was deduced within the 
double logarithmic approximation, i.e., $\alpha_s \ln{Q^2\over
\Lambda_{QCD}^2}\ln{1\over x} \sim 1$, with the result of
\begin{equation}
\left. {d\sigma_{\gamma^*_LN\rightarrow VN}\over dt} \right|_{t=0}~=~
  {4\pi^3 \Gamma_V M_V\over 3\alpha_{EM}Q^6}
  \,\eta^2_V
  \,\left|\alpha_s(Q^2)\left(1+i\beta\right)xG_N(x,Q^2)\right|^2
\ .
\label{eq0a}
\end{equation}
Here, $\Gamma_V$ stands for the decay width of the vector meson into an
$e^+e^-$ pair, $\beta={\hbox{Re}{\cal A}\over\hbox{Im}{\cal A}} \approx
{\pi\over 2}
{\partial\ln[Im A]\over\partial \ln x}$ 
is the relative contribution 
of the amplitude's real part, and the leading twist correction 
\begin{equation}
\eta_V~\equiv~{1\over 2}{\int\!{dz\over z(1-z)}\int\!d^2k_t\,\phi_V(z,k_t)
                    \over\int\!dz\int\!d^2k_t\,\phi_V(z,k_t)} \ ,
\label{eq0b}
\end{equation}
accounts for the difference between the vector meson's decay into an $e^+e^-$ 
pair and diffractive vector meson production. 
Here $\phi_V(z,k_t)$ is the wave function of the longitudinally 
polarized vector meson.  We implicitly use the light-cone gauge which provides for 
an unambigous separation of the 
$k_t$ dependence of the meson (photon) wave function 
and gluon degrees of freedom. 
Note that the original formula deduced in Ref. \cite{Brod94} lacks a 
factor of 4. This misprint has been corrected in Ref. \cite{hepph}.
In Ref. \cite{hepph}, it was shown that the formula in Eq.
(\ref{eq0a}) is valid also within the more  conventional leading $\alpha_s
\ln{Q^2\over\Lambda_{QCD}^2}$ approximation. 
Although, in principle, hard diffractive processes are expressed in terms of
non-diagonal parton densities, an analysis \cite{Freund} of the QCD
evolution equations for non-diagonal parton distributions at small $x$
shows that the difference between the diagonal and non-diagonal parton
distributions is small in the kinematic region discussed in this
paper.

In Ref. \cite{hepph},
also next-to-leading order (NLO) as well as higher twist corrections 
were introduced. Firstly, it was argued that the strong coupling
constant and the nucleon's gluon density have to be evaluated not at
$Q^2$ but at a $Q^2_{eff}$.  This is due to the so-called ``rescaling of 
hard processes'' which will be discussed in more depth later.  And, secondly,
a suppression factor $T(Q^2)$ was deduced which measures the deviation
of the cross section from its asymptotic prediction in Eq. (\ref{eq0a}), and
which stems from the transverse Fermi motion of the quarks in the produced
vector meson.  This yields \cite{hepph} 
\begin{equation}
\left. {d\sigma_{\gamma^*_LN\rightarrow VN}\over dt} \right|_{t=0}~=~
  {4\pi^3 \Gamma_V M_V\over 3\alpha_{EM}Q^6}
  \,\eta^2_V\,T(Q^2)
  \,\left|\alpha_s(Q_{eff}^2)\left(1+i\beta\right)xG_N(x,Q_{eff}^2)\right|^2
\ ,
\label{eq0c}
\end{equation}
with the correction factor
\begin{equation}
T(Q^2)~=~\left[{Q^4\over 4}~
\int\!{dz} \int\!d^2k_t~\phi_V(z,k_t)~\Delta_t~\phi_\gamma(z,k_t)
\over {\int\!{dz\over z(1-z)} \int\!d^2k_t~\phi_V(z,k_t)}  \right]^2 
\ ,
\label{eq0d}
\end{equation}                                                
where
\begin{equation}
\phi_{\gamma}(z,k_t)={1\over Q^2+{k_t^2+m^2\over z(1-z)}}
\label{eq0e}
\end{equation}
is the photon's $q\bar q$ light-cone wave function, $\Delta_t$ is the 
transverse Laplacian
$\Delta_t=\Sigma(d/dk_i)^2$,
and where, for the production of light mesons, the 
current quark mass was set to zero. 

In this investigation, we focus on the photo- and electroproduction of
heavy quarkonium ($J/\psi$ and $\Upsilon$), and we extend the respective 
formalism to the production of transversely polarized heavy
vector mesons as well.
Note that, for sufficiently heavy quark mass, the
applicability of the QCD factorization theorem to the 
diffractive photoproduction of {\em transversely} polarized vector mesons 
can be justified because the transverse size of quarkonium decreases 
with the mass of the heavy quarks.
The result for the forward differential cross section for photo- and 
electroproduction of heavy vector mesons, which will be deduced in detail 
in the following, is
\begin{eqnarray}
\left. {d\sigma_{\gamma^{(*)}N\rightarrow VN}\over dt} \right|_{t=0}~=~
  {4\pi^3 \Gamma_V M_V^3\over 3\alpha_{EM}(Q^2+4m^2)^4}
  &&\,\eta^2_V\,T(Q^2)
  \,\left|\alpha_s(Q_{eff}^2)\left(1+i\beta\right)xG_N(x,Q_{eff}^2)\right|^2
\nonumber
\\
&&\times\left(R(Q^2)+\epsilon {Q^2\over M_V^2}\right) 
\ .
\label{eq0f}
\end{eqnarray}
Here, $\eta_V$ is again the leading twist correction of Eq. (\ref{eq0b}),
the factor $T(Q^2)$, which was introduced in
Ref. \cite{hepph}, accounts for effects related to the quark motion in the 
produced vector meson, and 
$\epsilon = {1-y\over 1-y+y^2/2}$ is a parameter related to the (virtual) 
photon's polarization.
Here, $y$ is the energy fraction (in the target rest frame)
transferred from the scattered lepton to the target.  A value of $\epsilon=0$
corresponds to purely transverse polarization -- which is always the case
for real photons, i.e., $Q^2 = 0$ -- and $\epsilon = 1$ refers to an equal mix
of longitudinal and transverse polarizations.  The latter is typical for HERA
kinematics at large $Q^2$.
The  factor $R(Q^2)$ parameterizes the relative contribution of the 
production of transversely polarized vector mesons as compared to the naive 
prediction, i.e., ${\sigma_T\over \sigma_L} = R(Q^2)\,{M_V^2 \over Q^2}$ 
instead of simply ${\sigma_T\over \sigma_L} = {M_V^2\over Q^2}$.  

In difference from Ref.\cite{hepph}, the current quark mass was not set to 
zero, i.e., we kept leading powers over ${1 \over Q^2+4m^2}$ and not just 
$1\over Q^2$.  This, in turn, yields for the correction factors $T(Q^2)$ and 
$R(Q^2)$:
\begin{eqnarray}
T(Q^2)&=&\left[{\left(Q^2+4m^2\right)^2\over 4}~
{\int\!dz\
\int\!d^2k_t~\phi_V(z,k_t)~\Delta_t~\phi_\gamma(z,k_t)
    \over
 \int\!{dz\over z(1-z)}\int\!d^2k_t~\phi_V(z,k_t)}   \right]^2
\ ,
\label{eq0g}
\\
R(Q^2)&=&\left[{m^2 \over 4M_V^2}~
 {\int\!{dz\over z^2(1-z)^2}
        \int\!d^2k_t~\phi_V(z,k_t)~\Delta_t~\phi_\gamma(z,k_t)
    \over
 \int\!dz
\int\!d^2k_t~\phi_V(z,k_t)~\Delta_t~\phi_\gamma(z,k_t)}        \right]^2
\ ,
\label{eq0h}
\end{eqnarray}
where we employed again $\phi_\gamma(z,k_t)$ of Eq. (\ref{eq0e}).

The $T(Q^2)$ and $R(Q^2)$ displayed in the above constitute one of our main
original new results. 
These formulae are derived by building a decomposition over 
the {\em transverse} distance between the bare quarks, i.e., over 
powers of ${1\over Q^2+4m^2}$.  However, some caution is necessary at
this point.
The accuracy of this approximation for the calculation of $R$ at 
$Q^2 \gg M_V^2$ can be questioned because of an enhancement of end point 
($z=0$ and $z=1$) contributions at large $Q^2$.  But in these 
kinematics, the production of longitudinally polarized vector meson 
would dominate \cite{Brod94}.
In order to be able to evaluate the correction factors of Eqs. (\ref{eq0g}) and
(\ref{eq0h}), we need the 
light-cone wave function of the $q\bar q$ leading Fock state 
in the vector meson.  We will discuss this quantity in detail in the next
section.

Our master formula in Eq. (\ref{eq0f}) yields a few fundamental predictions:
1) the cross sections raise with energy very rapidly due to the
presence of the gluon density which increases fast
at small $x$, 2) the $t$-slope is expected to be almost the same for all hard 
diffractive processes of the kind studied 
here\footnote{This is because the $t$-slope in the hard vertex scales like the
maximum of the mass of the heavy quark and $Q^2$.}, 
and 3) the production
of longitudinally polarized vector mesons will dominate at large $Q^2$.
Note, also, that this is only a leading order analysis, and to achieve
a non-ambiguous interpretation of the processes considered here it
would be necessary to evaluate also
more accurately NLO corrections as well as the higher twist
 the contribution of the $|q\bar qG\rangle$ 
component in the light-cone wave functions of the photon and the produced
vector meson.

\subsection{The color-dipole cross section}
\label{II.B}

As discussed at length in Refs. \cite{Brod94}, \cite{Collins} and 
\cite{hepph}, due to the QCD factorization theorem and the large 
longitudinal coherence length, 
$l_c \approx {1\over 2 m_Nx}$,
associated with high energy (small $x$) diffractive processes,
in leading order in $\alpha_s \ln{Q^2\over\Lambda_{QCD}^2}$, the 
amplitude for hard diffractive vector meson production off a 
nucleon, depicted in Fig. 1,
can be written as a product of three factors,
\begin{equation}
{\cal A}_{\gamma^{(*)}N\rightarrow VN}~\propto~
   \Psi(\gamma^*\rightarrow q\bar q) \cdot
   \sigma_{q\bar q N} \cdot
   \Psi(q\bar q \rightarrow V)
\label{eq1}
\end{equation}
where $\Psi(\gamma^{(*)}\rightarrow q\bar q)$ is the light-cone wave function
for a photon to split into a $q\bar q$ pair, $\sigma_{q\bar q N}$ is the
interaction cross section of the $q\bar q$ pair with the target nucleon,
and $\Psi(q\bar q \rightarrow V)$ is the amplitude for the $q\bar q$ pair
to transform into the vector meson $V$ in the exit channel.  

\figa{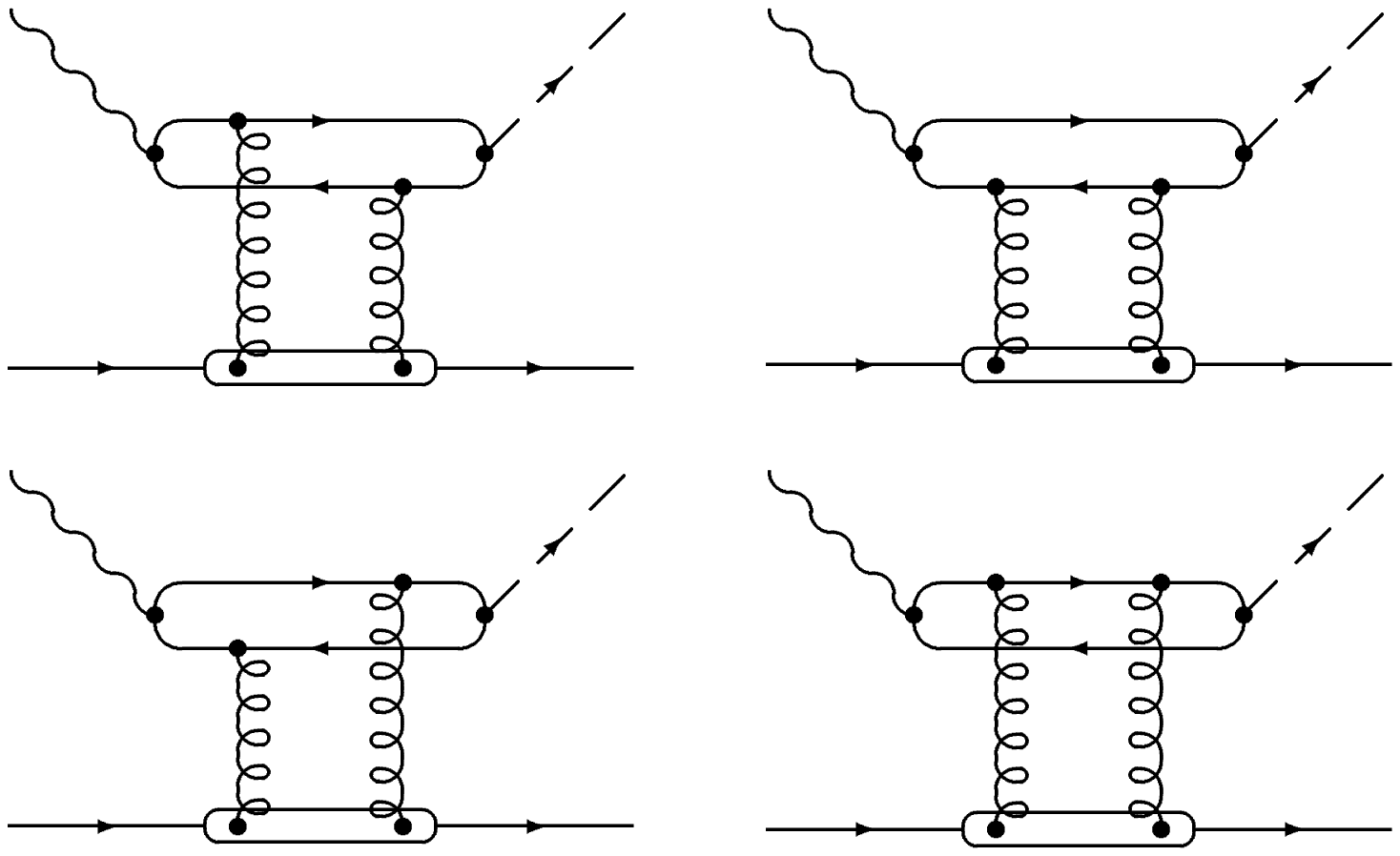}{5.0}
{Feynman diagrams relevant for the evaluation of the amplitude
for diffractive production of vector mesons, i.e., the $\gamma^{(*)}+N 
\rightarrow V + N$ process, in leading $\alpha_s\ln{Q^2\over\Lambda_{QCD}^2}$ 
approximation.}

As was shown in Ref. \cite{hepph}, for sufficiently large $Q^2$ and 
longitudinal polarization, the above process is dominated by $q\bar q$ 
configurations where the quark and antiquark are separated by a small 
transverse distance $b$.  Then, $\sigma_{q\bar q N}$ is the color-dipole cross 
section \cite{BBFS93,FMS93}
\begin{equation}
\sigma_{q\bar q N}(x,b)~=~{\pi^2\over 3}~b^2\,\left[\alpha_s(Q_{eff}^2)\,
xG_N(x,Q_{eff}^2)\right]_{x={Q^2+M_V^2\over s},
                        ~~Q_{eff}^2={\lambda\over b^2}} 
\ .
\label{eq2}
\end{equation}
Qualitatively, Eq. (\ref{eq2}) can be understood in the following way:
The four diagrams of Fig. 1 lead to an expression in the amplitude of the form 
\begin{equation}
\sigma_{\gamma^* N}
~\propto~\int\!dz\int\!d^2l_t~\ldots~
     [2\phi_\gamma(z,k_t)-\phi_\gamma(z,k_t+l_t)-\phi_\gamma(z,k_t-l_t)] \ ,
\label{eq3}
\end{equation}
where the Sudakov variable $z$ denotes the fraction of the photon's momentum 
carried by one of the quarks, $\pm k_t$ is their transverse momentum,
and $l_t$ is the gluons' transverse momentum.  For small $l_t$, this yields 
\begin{equation}
\sigma_{\gamma^* N}
~\propto~\int\!dz d^2 l_t~~\ldots~l_t^2\Delta_t \phi_\gamma(z,k_t) \ . 
\label{eq4}
\end{equation}
Via Fourier transform into the transverse impact parameter space
and after pulling out the wave function of the $\gamma^*$, we obtain:
\begin{equation}
\sigma_{q\bar q N}~\propto~b^2 \ .
\label{eq5}
\end{equation}  
The gluon density, $xG_N$, arises as the diagrams in Fig. 1 represent not 
simple two-gluon exchange but rather the coupling to the full non-perturbative 
gluon ladder.  For further details see Ref. \cite{rad} where the quantity 
$\sigma_{q\bar q N}$ was derived rigorously.  In the following, we will show 
that -- due to the large value of the current quark mass -- the dominance of 
short distances holds for diffractive production of heavy flavors also for 
$Q^2=0$ and both for longitudinal as well as transverse polarizations.  

Note that, due to the difference in the invariant mass between the
photon and the vector meson, the light-cone momentum fractions of the gluons
in the initial and final state, $\beta_i$ and $\beta_f$, are not the 
same, and therefore, in principle, an off-diagonal gluon distribution should 
enter into Eqs. (\ref{eq0f}) and (\ref{eq2}).  This was first recognized in
Ref. \cite{hepph}, and then elaborated on in Refs. \cite{radyu}, 
\cite{hood} and \cite{Freund}.
A simple kinematical consideration yields
$\beta_i \approx {M_X^2+\langle l_t^2\rangle +Q^2 \over Q^2+M_V^2}\,x$ and
$\beta_f \approx {M_X^2+\langle l_t^2\rangle -M_V^2 \over Q^2+M_V^2}\,x$,
where $\langle l_t^2 \rangle$ is the average transverse momentum of the 
exchanged gluons and $M_X^2=\langle {k_t^2+m^2\over z(1-z)} \rangle$ 
is the invariant squared mass of the produced $q\bar q$ pair.
Within the $\alpha_s\ln{Q^2\over \Lambda_{QCD}^2}$ approximation,
the non-diagonal gluon distribution is 
shown \cite{Freund} to be not far -- at the small $x$ that are important
experimentally --
from the diagonal one.  This is because, within this approximation, the
appropriate energy denominators only weakly depend on $\beta_i$.

\subsection{Rescaling of hard processes}
\label{II.C}

As outlined in detail in Ref. \cite{hepph}, the parameter $\lambda$, which 
fixes the scale in the gluon density and the strong coupling in Eqs. 
(\ref{eq0f}) and (\ref{eq2}), is determined by comparison with the 
longitudinal structure function, $F_L(x,Q^2) \propto \left.yG_N(y,Q^2)
\right|_{y\,\approx\,2.5x}$, i.e., by setting 
\begin{equation}
\left.yG_N(y,Q^2)\right|_{y\,\approx\,2.5x}
\propto \int\!d^2b~dz~|\phi_{\gamma_L^*}(z,b)|^2~\sigma_{q \bar q N}(2.5x,b)
\ ,
\label{eq6}
\end{equation}
where 
\begin{equation}
\phi_{\gamma_L^*}(z,b)~=~2Q\,z(1-z)\,K_0\!\left(b\sqrt{Q^2z(1-z)+m^2}\right)
\label{eq6a}
\end{equation}
is the light-cone wave function of the $q\bar q$ leading Fock component in a
longitudinally polarized virtual photon, and $m$ is the current quark mass,
which was set to zero when we evaluated Eq. (\ref{eq6}) (since the 
contribution of charm quarks to $F_L$ is small in the considered kinematics).  
In  Eqs. (\ref{eq6}) 
and (\ref{eq6a}), $b$ is the transverse distance between the quark and 
antiquark within the photon.  The quantity $\lambda$ is adjusted such that the 
average $b=b_{\sigma_L}$, which dominates the integral on the right hand side 
of Eq. (\ref{eq6}), is related to $Q^2$ just via the equality 
$b_{\sigma_L}^2 ={\lambda \over Q^2}$.  In other words, for the 
longitudinal structure function,
the virtuality that corresponds to the dominant transverse distance 
$b_{\sigma_L}$ is just the virtuality of the process.  This yields $\lambda 
\sim 8.5$ for $x=10^{-3}$.

In the same fashion, we can now rewrite the amplitude for diffractive vector
meson production as
\begin{equation}
{\cal A}_{\gamma^*_LN\rightarrow VN}~\propto~
\alpha_s(Q_{eff}^2)\,xG_N(x,Q_{eff}^2)
\int\!dz\,d^2b~\phi_{\gamma^*_L}(z,b)~b^2~\phi_V(z,b)
\ ,
\label{eq7}
\end{equation}
where we pulled the gluon density at an average $b=b_V$ out of the integral,
i.e., $Q_{eff}^2 \sim {\lambda\over b_V^2}$, and with the $q\bar q$
leading  Fock state light-cone wave function of the 
vector meson, $\phi_V(z,b)$.
In Fig. 2, we show $b_V$ and $Q_{eff}^2$ for the longitudinal structure
function as well as for diffractive production of (longitudinally polarized)
$\rho^o$, $J/\psi$ and $\Upsilon$ mesons.  The wave functions $\phi_V(z,b)$
that were used to evaluate Eq. (\ref{eq7}) will be discussed in more detail
later.

\figa{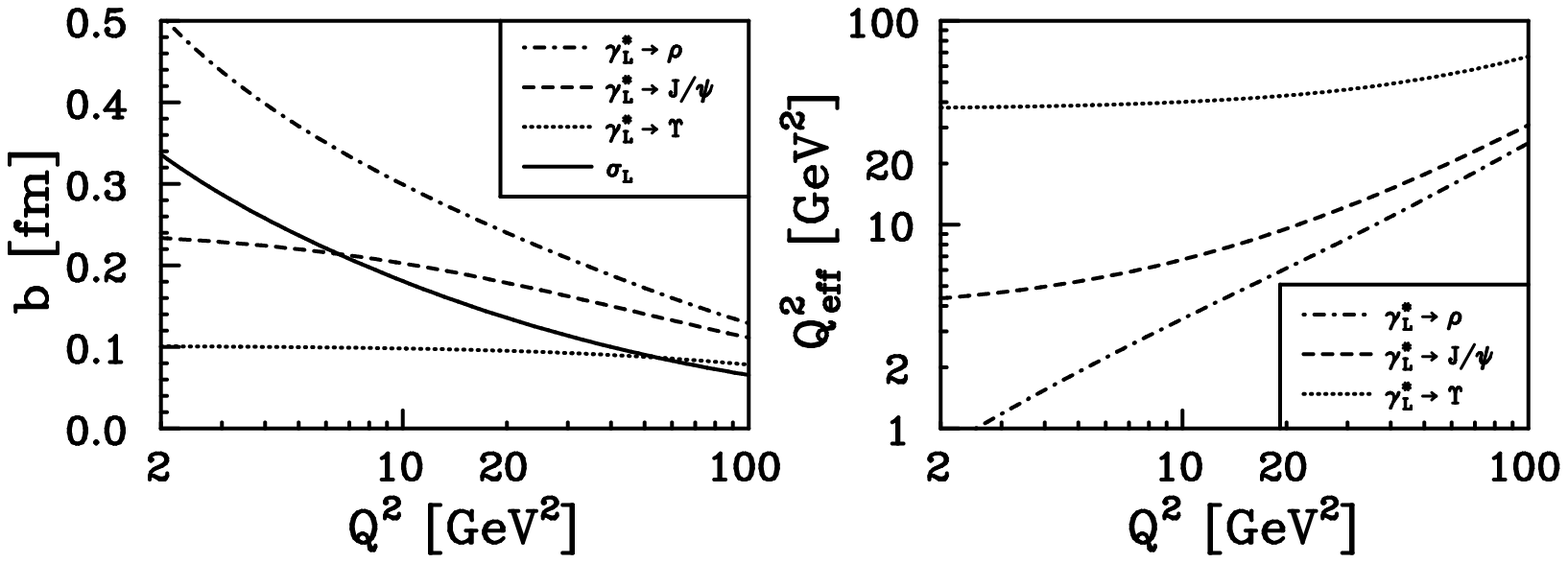}{6.5}
{Average transverse distances effective in the evaluation of 
the longitudinal structure function as well as for diffractive production of 
longitudinally polarized $\rho^o$, $J/\psi$ and $\Upsilon$ mesons.  Also shown 
are the resulting effective scales, $Q_{eff}^2$, for diffractive vector meson 
production.}

It can be seen from Fig. 2 that the relevant transverse distances for $\rho^o$
electroproduction are larger than those characteristic for the longitudinal 
structure function, i.e., $b_\rho(Q^2) > b_{\sigma_L}(Q^2)$.  Therefore, for 
$\rho^o$ production, the virtuality $Q^2_{eff}$ that enters in the argument of 
$\alpha_s(x,Q_{eff}^2)\,xG_N(x,Q_{eff}^2)$ is smaller than $Q^2$.  We find, to
leading order, 
\begin{equation}
b_V(Q^2)~\approx~b_{\sigma_L}(Q_{eff}^2)
\ ,
\label{eq8}
\end{equation}
which, for $\rho^o$ production, yields $Q_{eff}^2 \approx Q^2\left({b_{\sigma_L}(Q^2)
\over b_V(Q^2)}\right)^2$.
Our Eq. (\ref{eq8}) is an approximate relation designed to overcome the scale 
ambiguity which is inherent to leading order calculations.  This ``rescaling of
hard processes" effectively relates the scales in different processes via the 
dominant $q \bar q$ distances in the respective quark loops.  We termed this 
``$Q^2$ rescaling" in Ref. \cite{hepph}.  The difference between $Q^2_{eff}$ 
and $Q^2$ indicates that substantial next-to-leading order corrections should 
be present in those processes.  Applying the same method to $J/\psi$ and
$\Upsilon$ production yields a $Q^2_{eff}$ which is significantly larger than 
the estimate ${\overline Q}^2={Q^2+M_V^2\over 4}$ of Refs.
\cite{Ryskin,Ryskin2}. 
Fig. 2 also indicates that the relevant transverse distances are
small, and hence 
the QCD factorization theorem
is applicable, for $\rho^o$ production at large $Q^2$ and heavy 
meson photo- and electroproduction.

\subsection{Production of transversely polarized vector mesons}
\label{II.D}

The discussion in the above refers to the production of longitudinally 
polarized vector mesons only.  For light vector mesons, the formalism at hand 
cannot be extended to transverse polarizations because of 
the endpoint singularities, i.e., the contribution from very asymmetric 
$q\bar q$ pairs with $z \sim 0$ or $1$, where non-perturbative 
effects dominate. For 
the production of heavy quarkonia $Q\bar Q$, when $\alpha_s(M_Q^2)\ll 1$
and ${q_o \over 4M_Q^2} \gg r_T$ however, effects of large transverse 
distances are strongly suppressed. Here, $r_T$ is the radius of the hadron
target. So, the production of transversely polarized heavy
quarkonia can be legitimately evaluated using the QCD factorization
theorem.  At the same time, for $Q^2 \gg M_V^2$ the end point contribution  
($z \sim 0$ or $1$) is enhanced
in the amplitude for the diffractive elctroproduction of 
transversely polarized vector mesons.  Thus, the region of applicability of 
non-relativistic wave function models for heavy quarkonia 
to the production of transversely polarized vector mesons
(but not of the applicability of the QCD factorization theorem)
is restricted by the kinematical constraint $Q^2 \leq M_V^2$.

Employing the notations of Ref. \cite{Brod94}, 
the wave functions of longitudinally and transversely polarized
photons and heavy vector mesons 
can
be expressed as
\begin{eqnarray}
\phi_{\gamma_L}^{\lambda_1\lambda_2} &=& 
2Q\,z(1-z)\,\phi_\gamma(z,b)\,\delta^{\lambda_1}_{-\lambda_2}
\ ,
\label{eq_ref1}
\\
\phi_{\gamma_T}^{\lambda_1\lambda_2} &=& 
m \pmatrix{ \mp 1 \cr -i \cr}\phi_\gamma(z,b)\,\delta^{\lambda_1}_{\lambda_2}
   ~+~
   \pmatrix{ i(2z-1){\hat b}_x \mp {\hat b}_y \cr
            \pm {\hat b}_x + i(2z-1) {\hat b}_y \cr}
   {\partial\phi_\gamma(z,b)\over \partial b}\,\delta^{\lambda_1}_{-\lambda_2}
\ ,
\label{eq_ref2}
\\
\phi_{V_L}^{\lambda_1\lambda_2} &=& 
 -2M_V\,\phi_V(z,b)\,\delta^{\lambda_1}_{-\lambda_2}
\ ,
\label{eq_ref3}
\\
\phi_{V_T}^{\lambda_1\lambda_2} &=& 
{m \over z(1-z)} \pmatrix{ \mp 1 \cr -i \cr}
\,\phi_V(z,b)\,\delta^{\lambda_1}_{\lambda_2}
\ ,
\label{eq_ref4}
\end{eqnarray}
where 
\begin{equation}
\phi_\gamma(z,b)~=~K_0\!\left(b\sqrt{Q^2z(1-z)+m^2}\right)
\label{eq10a}
\end{equation} 
and $\phi_V(z,b)$ refer to the $q\bar q$ light-cone wave functions of the 
photon and the heavy vector meson, respectively.  For the derivation 
of Eqs. (\ref{eq_ref3}) and (\ref{eq_ref4}) it was assumed, in line with the 
non-relativistic character of heavy quarkonium, that, in the center of mass 
system, the vector meson's wave function is a pure angular momentum $L=0$ 
state.  This selects spin $S=0$ (or helicities $\lambda_2=-\lambda_1$) for 
the longitudinal polarization and $S=1$ (or $\lambda_2=\lambda_1$) for the
transverse polarizations, with the same spatial wave function $\phi_V(z,b)$.  
Here, $\lambda_{1,2}$ are the helicities of the quark and antiquark, 
respectively. 
For transverse polarization, 
the restriction through the wave function of heavy quarkonia selects the 
component in the wave function of the virtual photon
which is proportional to the mass of the heavy quark.
This is just opposite to the production of mesons
built of light quarks where this component in the photon's wave 
function is negligible \cite{Mueller5}.

This gives for the kernels of the longitudinal and transverse amplitudes:
\begin{eqnarray}
V_L(z,b) &~=~ {1\over 2}\sum_{\lambda_1\lambda_2}\,
         {\phi_{\gamma_L}^{\lambda_1\lambda_2}}^\dagger\,
         \phi_{V_L     }^{\lambda_1\lambda_2}
    &~=~ -4QM_V\,z(1-z)\,\phi_\gamma(z,b)\,\phi_V(z,b)
\ ,
\label{eq9}
\\
V_T(z,b) &~=~ {1\over 4}\sum_{\lambda_1\lambda_2}\,
       {\phi_{\gamma_T}^{\lambda_1\lambda_2}}^\dagger\,
         \phi_{V_T     }^{\lambda_1\lambda_2}
    &~=~ {m^2\over z(1-z)}\,\phi_\gamma(z,b)\,\phi_V(z,b)
\ .
\label{eq10}
\end{eqnarray}




Note that in the limit $z\approx {1\over 2}$ and $M_V 
\approx 2m$, Eqs. (\ref{eq9}) and (\ref{eq10}) yield the naive prediction
${\sigma_L\over\sigma_T}=\left({V_L\over V_T}\right)^{\!2}
\approx{Q^2\over M_V^2}$
for the production ratios of longitudinal to transverse polarizations.
Also, due to the non-relativistic ansatz for the vector meson's wave function,
the spin structure of Eq. (\ref{eq10}) is such that there is no azimuthal 
asymmetry.  This is qualitatively different from diffractive two-jet production
in deep inelastic scattering \cite{wust}. 
Note, however, that in a fully relativistic description such an
azimuthal asymmetry would appear also for diffractive production 
of transversely polarized vector mesons due to the admixture 
of a $L=2$, $S=1$ component (to the standard $L=0$, $S=1$ 
state).
Note, also, that the non-relativistic approximation to the light-cone wave
function of transversely polarized heavy quarkonia becomes 
questionable for the diffractive electroproduction in the limit
$Q^2\gg M_V^2$.  This is because, in this kinematics, the end point
contributions $z=0$ and $z=1$ are enhanced.  But in QCD, in variance 
from non-relativistic quarkonium models, the wave function at asymptotical
$Q^2$ should be such that
$V_T \propto z(1-z)$. Such a behavior follows from the analysis of
pQCD diagrams for the wave function of heavy quarkonia.  This complication is 
practically unimportant because, in this kinematics, the production
of longitudinally polarized heavy quarkonia dominates.

Putting everything together, the factor $T(Q^2)$, which accounts for effects 
related to the quark motion in the produced vector meson, and the correction 
factor $R(Q^2)$, which parameterizes the relative contribution of the 
transverse production, can be written in transverse impact parameter space as
\begin{eqnarray}
T(Q^2)&=&\left[{\left(Q^2+4m^2\right)^2\over 4}~
       {\int\!dz\,z(1-z)\,\int\!db~\phi_V(z,b)~b^3~\phi_\gamma(z,b)
       \over
       \int\!{dz\over z(1-z)}~\phi_V(z,b=0)}            \right]^2
\ ,
\label{eq13}
\\
R(Q^2)&=&\left[{m^2 \over 4M_V^2}~
 {\int\!{dz\over z(1-z)}\int\!db~\phi_V(z,b)~b^3~\phi_\gamma(z,b)
  \over
  \int\!dz\,z(1-z)\int\!db~\phi_V(z,b)~b^3~\phi_\gamma(z,b)}      \right]^2
\ ,
\label{eq14}
\end{eqnarray}
where we used again $\phi_\gamma(z,b)$ of Eq. (\ref{eq10a}).  The $T(Q^2)$ and 
$R(Q^2)$, displayed in the above, are the leading expressions to order ${1\over
Q^2+4m^2}$, and they constitute our main original new results.  They are 
related to the quantities given in Eqs. (\ref{eq0g}) and (\ref{eq0h}) simply 
via a two-dimensional Fourier transformation.

\subsection{Leading twist expressions and comparison with other 
$k_t$-suppression estimates}
\label{II.E}

Note that the suppression factor $T(Q^2)$ of Eq. (\ref{eq13}) and the 
transverse to longitudinal production ratio $R(Q^2)$ of Eq. (\ref{eq14}) have 
contributions from leading and non-leading twist.  The corresponding leading 
twist expressions can be deduced by pulling the vector meson's wave function, 
$\phi_V(z,b)$, at $b=0$ out of the integral, i.e, by replacing $\phi_V(z,b)$ 
with $\phi_V(z,0)$.  The latter is equivalent to setting in the photon's 
wave function, $\phi_\gamma(z,k_t)$ of Eq. (\ref{eq0e}), $k_t$ to zero after 
differentiation, and it yields
\begin{eqnarray}
T_{LT}(Q^2)&=&\left[
             {\int\!{dz\over z(1-z)}
              \left({Q^2+4m^2 \over Q^2+{m^2\over z(1-z)}}\right)^{\!\!2}
              \phi_V(z,b=0)
             \over
              \int\!{dz\over z(1-z)}~\phi_V(z,b=0)}       \right]^2
\ ,
\label{eq14a}
\\
R_{LT}(Q^2)&=&\left[{m^2 \over 4M_V^2}~
             {\int\!{dz\over z^3(1-z)^3}
              \left({1\over Q^2+{m^2\over z(1-z)}}\right)^{\!\!2}\phi_V(z,b=0)
              \over
              \int\!{dz\over z(1-z)}
              \left({1\over Q^2+{m^2\over z(1-z)}}\right)^{\!\!2}\phi_V(z,b=0)}
                                                                   \right]^2
\ .
\label{eq14b}
\end{eqnarray}
This shows that, for these processes, a decomposition over twists is really 
an expansion in powers of $b^2$,
and ``leading twist" is equivalent to the $b \to 0$ limit, i.e., to considering
very small transverse distances (or ``pointlike hadrons") only.  
Specific to heavy quarkonium production is that, in addition to neglecting 
$k_t^2/Q^2$ and $m^2/Q^2$ terms as for light quarks, one also neglects 
terms of the form $k_t^2/m^2$.

Note that the expressions (\ref{eq14a}) and
(\ref{eq14b}) are stringent QCD predictions for heavy quark production
deduced in 
an expansion where $m$ is considered as a  large parameter.  
The leading term is proportional to the mass of the heavy quark in 
difference from
light quark production where the leading term is proportional to the
quark's transverse momentum. So, the formulae deduced in this paper
cannot be smoothly interpolated to the 
limit of the zero quark mass.

Furthermore, in the static limit of $m \rightarrow \infty$, which implies 
$\phi_V(z,k_t)=\delta\!\left(z-{1\over 2}\right)\,\phi_V(k_t)$ and $M_V = 2m$, 
the correction factors $T(Q^2)$, $R(Q^2)$, $T_{LT}(Q^2)$ and $R_{LT}(Q^2)$
reduce to 
\begin{eqnarray}
T(Q^2)~&\rightarrow&~1\,-\,32\,{\langle k_t^2 \rangle\over Q^2+4m^2}
\ , \label{eq14c} \\
R(Q^2)~&\rightarrow&~1
\ , \label{eq14d} \\
T_{LT}(Q^2)~&\rightarrow&~1
\ , \label{eq14e} \\
R_{LT}(Q^2)~&\rightarrow&~1
\ , \label{eq14f}
\end{eqnarray}
where
\begin{equation}
\langle k_t^2 \rangle~\equiv~{\int\!d^2k_t~k_t^2~\phi_V(k_t)
                              \over
                              \int\!d^2k_t~\phi_V(k_t)}
\ .
\label{eq14g}
\end{equation}

Recently, in two investigations \cite{Ryskin2,hood}, effects of the
transverse quark motion on diffractive charmonium production were discussed.
In Ref. \cite{hood}, the presence of a $Q^2$ independent (!) correction 
was claimed, which contradicts the strict asymptotic QCD 
result of \cite{Brod94}.
For photoproduction, the correction term of Ref. \cite{hood} is by a 
factor of 24 smaller than our leading twist, order ${\cal O}(k^2_t)$ 
correction of Eq. (\ref{eq14c}).  
To be able to compare with the result of Ref. \cite{Ryskin2}, we use the
expression of $T(Q^2)$ in transverse momentum space, i.e., Eq. (\ref{eq0g}).
The correction factor for $J/\psi$ photoproduction discussed in Ref. 
\cite{Ryskin2} can be obtained from our $T(Q^2=0)$ of Eq. (\ref{eq0g})
by approximating $\Delta_t\,\phi_\gamma(z,k_t)$ with the respective leading 
order expression in ${\cal O}\left(k_t^2/m^2\right)$ and by neglecting the 
longitudinal relative motion of the quarks, i.e., by setting $\phi_V(z,k_t)=
\delta\!\left(z-{1\over 2}\right)\,\phi_V(k_t)$.  In addition, a Gaussian form 
for the wave function $\phi_V(k_t)$ was assumed in Ref. \cite{Ryskin2}.  All 
of these approximations diminish the relative contribution of large quark momenta, 
and hence result in a significantly weaker suppression.  This was already 
pointed out in Ref. \cite{hepph} in a footnote.

\subsection{The $t$-slope of diffractive vector meson production}
\label{II.F}

It was demonstrated in Ref. 
\cite{Brod94} that, in the limits of fixed small $x$ and 
$Q^2 \rightarrow \infty$, the $t$-slope of the vector meson electroproduction 
cross section should be flavor independent and determined solely by the slope 
of the gluon-nucleon scattering amplitude.  However, the contribution of
finite $b\ne 0$ quark separations in the production amplitude of Eq. 
(\ref{eq7}) -- cf. Fig. 2 -- lead to a $Q^2$ and flavor dependence of the 
$t$-slope.  This effect can easily be incorporated into Eq. (\ref{eq7})
by evaluating the matrix element of the factor
$e^{-i z\vec q_t\cdot\vec b}+e^{iz\vec q_t\cdot\vec b} -
e^{-i (z\vec q_t+\vec l)\cdot\vec b} -  e^{i (z\vec q_t+\vec l)\cdot\vec b}$
between the wave functions of the photon and the vector meson. Here,
$\vec l$ is transverse momentum of one of the exchanged gluons, $\vec q_t$ is the 
transverse component of 
the four-momentum transfered to the target nucleon and $t=-|\vec q_t|^2$
(we neglect here terms proportional to $x$).
This amplitude can be written, in factorized  form, as a 
convolution integral over $l_t$ of the hard blob and the non-diagonal gluon 
distribution in the target at a virtuality $l^2$.
Since $b$ is small, it is reasonable to decompose this expression over 
powers of $b$ up to $b^2$. After that, the respective amplitude factorizes 
into a product of the hard blob (accounting for the $t$ dependence) and the 
non-diagonal gluon distribution in the target at virtuality $\sim {\lambda \over b^2}$.
Note that since the momentum  integrals in the hard blob depend on $b^2$
only logarithmically, the $b$ dependent term in the slope of the
gluon-nucleon amplitude should decrease with $b$ at least as
$b^2/\ln {b\over b_0}$.  So we can neglect it, to a first 
approximation, as compared to the effects of the form factor 
in the $\gamma_L^* \to V$ vertex, which are proportional to $b^2$.  Besides, 
studies of soft elastic scattering indicate that even for such processes 
the main contribution to the $t$-dependence of the amplitude 
comes from hadron form factors (if the energy is not so large 
that Gribov diffusion contributes).  Hence, we expect 
that, in the hard regime, the $b$ dependent term in the amplitude
will have a numerically small coefficient, in addition to being 
suppressed by the $\ln {b \over b_0} $ factor.

Thus, effectively, one should include, similar as for a form factor of 
a $Q \bar Q$ bound state \cite{Kogut}, in the integral on the right hand 
side of Eq. (\ref{eq7}) an additional factor of $e^{-i z\vec q_t\cdot\vec b}$,
where $\vec q_t$ is the three-momentum transfered 
to the target nucleon and $t=-|\vec q_t|^2$.  
Parameterizing as usual ${d \sigma \over dt}=A e^{B_Vt}$ for small
$t$, we can calculate the $Q^2$ dependence of 
$\Delta B_V(Q^2) \equiv B_V(Q^2) - B_V(Q^2 \rightarrow
\infty)$, i.e., the contribution of the hard blob of Fig. 1 to the 
$t$-dependence of the cross section, from
\begin{equation}
\Delta B_V(Q^2)~=~{1\over 2}\,
{\int\!dz\,d^2b~\phi_{\gamma_L^*}(z,b)~\sigma_{q\bar q N}(x,b)~\phi_V(z,b)
 ~z^2\,b^2 \over \int\!dz\,d^2b~\phi_{\gamma_L^*}(z,b)~
\sigma_{q\bar q N}(x,b)~\phi_V(z,b)    }
\ ,
\end{equation}
with the color dipole cross section $\sigma_{q\bar q N}(x,b)$ of Eq. 
(\ref{eq2}), and $\phi_{\gamma_L^*}(z,b)$ of Eq. (\ref{eq6a}) 
and $\phi_V(z,b)$, the $q\bar q$ light-cone wave functions of 
the photon and the vector meson, respectively.  
Results of such a calculation are presented 
in Fig. 3 for $J/\Psi$ and $\rho$-meson production.
Thus, the dependence of the cross section on $t$ contains information on the
distribution of color in the produced vector mesons.
Note that the 
experimentally observed $t$-slope for $J/\Psi$ photo- and electroproduction
and $\rho^o$ electroproduction at large $Q^2$ is of the order of
$B_V \approx 4 - 5$ GeV$^{-2}$.  The main conclusion from Fig. 3 is thus
that the $t$-slope of diffractive vector meson production is determined
mostly by the gluon-nucleon scattering amplitude, the differences in $B_V$
between different flavors are small for realistic $Q^2$, and they vanish in
the $Q^2\rightarrow\infty$ limit.

\figa{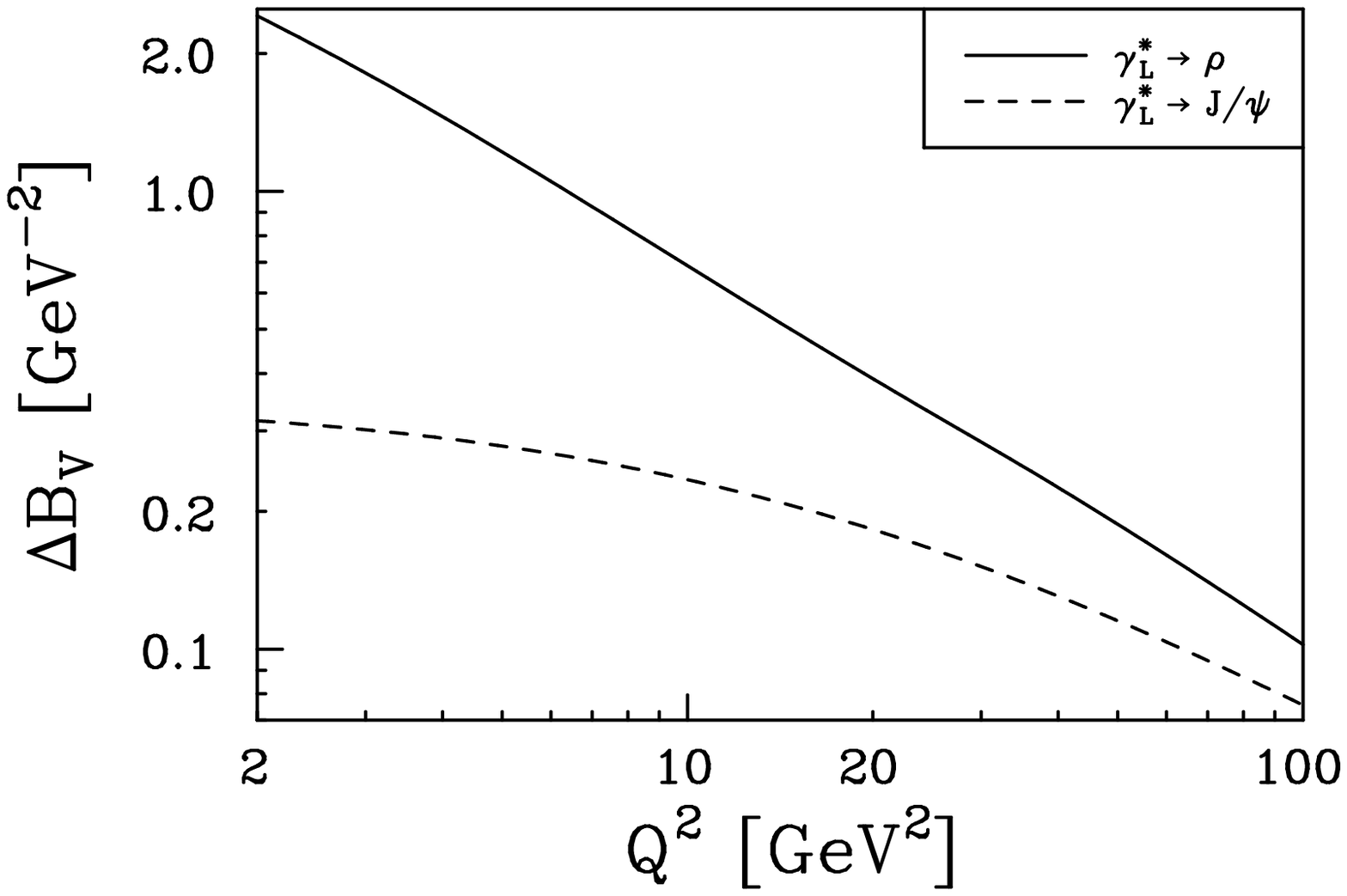}{4.0}
{The contribution of the hard blob to the $t$-slope of
diffractive electroproduction of longitudinally polarized $\rho^o$ and
$J/\psi$ vector mesons.}

We have demonstrated in Ref. \cite{hepph} that, at 
sufficiently small $x$, i.e.,
close enough to the low $x$-range probed at HERA, higher twist effects may 
become important.  This would also lead to a break-down of the universality 
of the $t$-slope.  This effect can be estimated by including double scatterings
of the $q \bar q$ pair off the nucleon \cite{AFS,hepph,Frank96}.  Neglecting
the small difference of the average $b$ for single and double scattering,
we can calculate the $t$-slope of the rescattering amplitude from
\begin{equation}
\left.{d\sigma \over dt}\right|_{screen}=\left.{d\sigma \over dt}\right|_{t=0}
\,\cdot\,\left|\,e^{Bt/2}~-~{1 \over 16 \pi B}\,
 {\langle\sigma_{q\bar q N}^2(x,b)\rangle 
  \over 
  \langle\sigma_{q\bar q N}  (x,b)\rangle}
 \left(e^{Bt/4}\,+\,r\,e^{{\tilde B}t/4}\right)\,\right|^2.
\label{slope}
\end{equation}
Here $\left.{d\sigma \over dt}\right|_{t=0}$ is the cross section given in 
Eq. (\ref{eq0c}) and 
\begin{equation}
{\langle\sigma_{q\bar q N}^2(x,b)\rangle 
 \over 
 \langle\sigma_{q\bar q N}(x,b)\rangle}~=~
{\int\!dz\,d^2b~\phi_{\gamma^*_L}(z,b)~\sigma_{q\bar q N}^2(x,b)~\phi_V(z,b) 
 \over
 \int\!dz\,d^2b~\phi_{\gamma^*_L}(z,b)~\sigma_{q\bar q N}(x,b)~\phi_V(z,b)} 
\ ,
\label{sigma2}
\end{equation}
where $\sigma_{q\bar q N}(x,b)$ is again the color dipole cross section
of Eq. (\ref{eq2}).  The quantity denoted $r$ is the ratio of the inelastic to
elastic diffractive production of vector mesons at large $Q^2$, and 
experimentally $r \approx 0.2$.  $\tilde B$ is the slope of the inelastic 
production, i.e., the $\gamma^*_L+p \rightarrow V+X$ process, and it is, so 
far, not well known experimentally.  Since for this process, in 
difference from 
the elastic production, there is essentially no form factor at the nucleon 
vertex, the quantity $\tilde B$ is much smaller than the elastic slope $B$, and
a natural guess is ${\tilde B} \approx 1 - 2$ GeV$^{-2}$.  Experimentally
\cite{H1}, the ratio of inelastic to elastic $J/\Psi$ production is of the 
order of $0.5-0.7$, i.e., $ {r\,B \over {\tilde B}}\sim 0.5-0.7$.  

For our numerical estimates we set $r=0.25$, $B(x \sim 10^{-2})=5$ GeV$^{-2}$, 
and ${\tilde B}=2$ GeV$^{-2}$.  In Fig. 4a we show the $t$-dependence of the
diffractive $\rho^o$ electroproduction cross section, i.e., $\left.{d\sigma 
\over dt}\right|_{screen}/\left.{d\sigma \over dt}\right|_{t=0}$ of Eq. 
(\ref{slope}).  Since the color dipole cross section $\sigma_{q\bar q N}(x,b)$
of Eq. (\ref{eq2}) is proportional to $xG_N(x,\lambda/b^2)\propto x^{0.2-0.3}$,
Eq. (\ref{slope}) leads to an increase of the $t$-slope with decreasing
$x$.  This can be seen from Fig. 4a, where we compare $\left.{d\sigma \over 
dt}\right|_{screen}$ for $x=10^{-2}$ (dashed line) and $x=10^{-4}$ (solid line)
with the leading twist result $e^{-Bt}$ (dotted line).

\figa{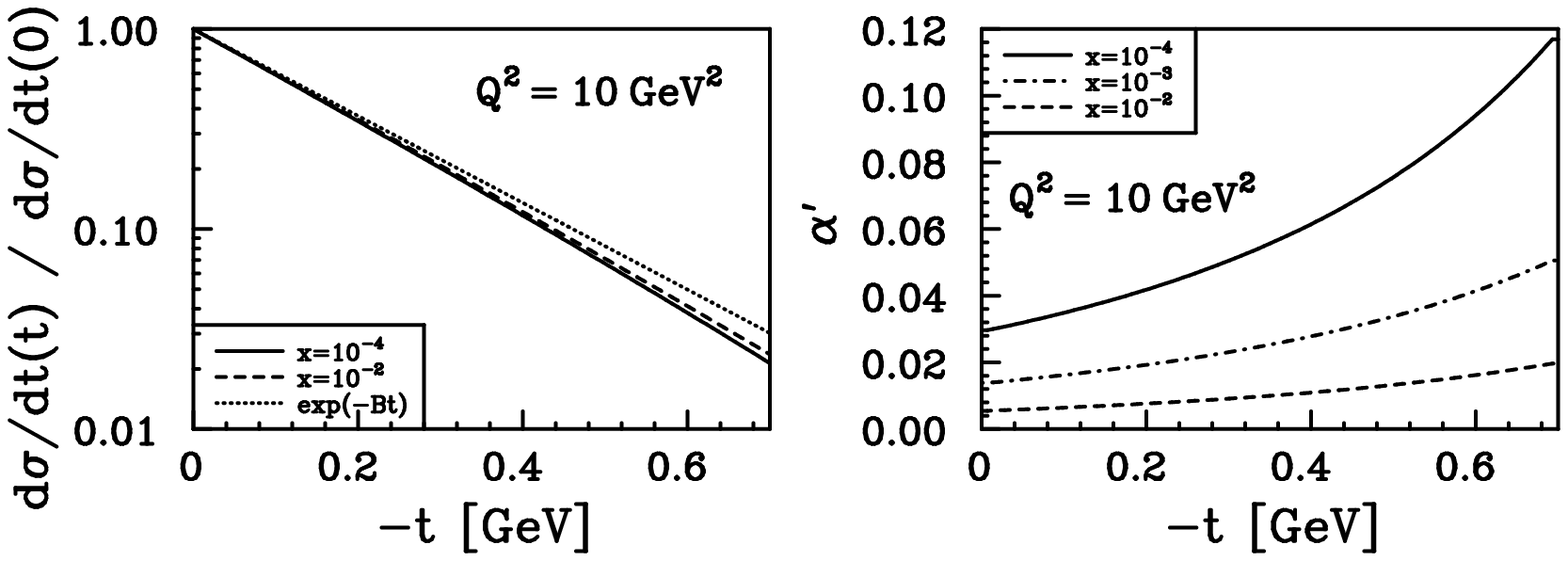}{6.5}
{The $t$-dependence of the diffractive vector meson
production cross section, i.e., $\left.{d\sigma \over 
dt}\right|_{screen}/\left.{d\sigma \over dt}\right|_{t=0}$ of Eq.
(\protect\ref{slope}), and the change of the $t$-slope with energy,
i.e., $\alpha^{\prime}(t)$ of Eq. (\protect\ref{alprim}).  Results are
shown for $\rho^o$ electroproduction at $Q^2=10$ GeV$^2$.}

The change of the $t$-slope with ``energy" $s={Q^2\over x}$ is usually 
parameterized in the form
\begin{equation}
B(s)~=~B(s_0)~+~2\,\alpha^{\prime}(t)\,\ln\left[{s\over s_0}\right]
\ ,
\label{alprim}
\end{equation}
and the quantity $\alpha^{\prime}$ increases with $-t$.  This can be seen
from Fig. 4b, where we show $\alpha^{\prime}(t)$ as a function of $t$ for
various $x$.  Again, due to the increase of the color dipole cross section
$\sigma_{q\bar q N}(x,b)$ with energy, the increase of $\alpha^{\prime}$
with $-t$ is more dramatic for smaller $x$ (larger energies).  Note that
the numerical results shown in Fig. 4 refer to $\rho^o$ electroproduction
at $Q^2=10$ GeV$^2$.  As can be seen from Fig. 2, the respective $Q^2_{eff}$
is very similar to the $Q^2_{eff}$ relevant for $J/\psi$ photoproduction,
and hence the numerical estimates shown in Fig. 4 should thus be approximately
valid also for $J/\psi$ photoproduction.  Fig. 4 suggests that a study of the
$t$-slopes of diffractive vector meson production may yet provide another 
sensitive probe of the dynamics of hard diffraction.

\section{The quarkonium light-cone wave function}
\label{III}

In order to be able to evaluate the asymptotic correction $\eta_V$ of Eq. 
(\ref{eq0b}) as well as the $T(Q^2)$ and $R(Q^2)$ of Eqs. (\ref{eq0g}) and
(\ref{eq0h}) or (\ref{eq13}) and
(\ref{eq14}), we need the light-cone wave function of the 
$q\bar q$ leading Fock state in the vector meson.  We will discuss this 
quantity in detail in this section.  Note also that, as a result of the 
factorization theorem in QCD, it is the distribution of bare (current) quarks 
that enter in the description of these hard processes, and therefore, a priori,
there should be no simple relation between this quantity and 
non-relativistic potential models.

\subsection{Non-relativistic potential models}
\label{III.A}

Due to the large value of the quark mass, it is generally assumed that
a non-relativistic ansatz with a Schroedinger equation and an effective
confining potential yields a fairly good description of heavy quarkonium
bound states.  The various models -- see Ref. \cite{eichten} for an overview --
differ in the functional form of the potential, but they all give a reasonable 
account of the $c\bar c$ and $b\bar b$ bound state spectra and decay widths.  
The same holds for the light-front QCD bound state calculation of Ref.
\cite{perry}.  In Fig. 5, we display the quantities $R_{00}(r)$ (normalized 
such 
that $\int\!dr\,r^2\,|R_{00}(r)|^2=1$) and $4 \pi^2 k^2 \phi_V(k)$ (normalized
such that ${1\over (2\pi)^3}\int\!d^3k\,\left|\phi_V(k)\right|^2=1$).  For
the latter, we also plot a Gaussian fit adjusted to reproduce $\phi_V(k)$ at 
small $k$.  It turns out that the wave functions can be well approximated at 
small $k$ by Gaussians, while, at large $k$, they fall off much slower and
they display a significant high momentum tail. 
\figa{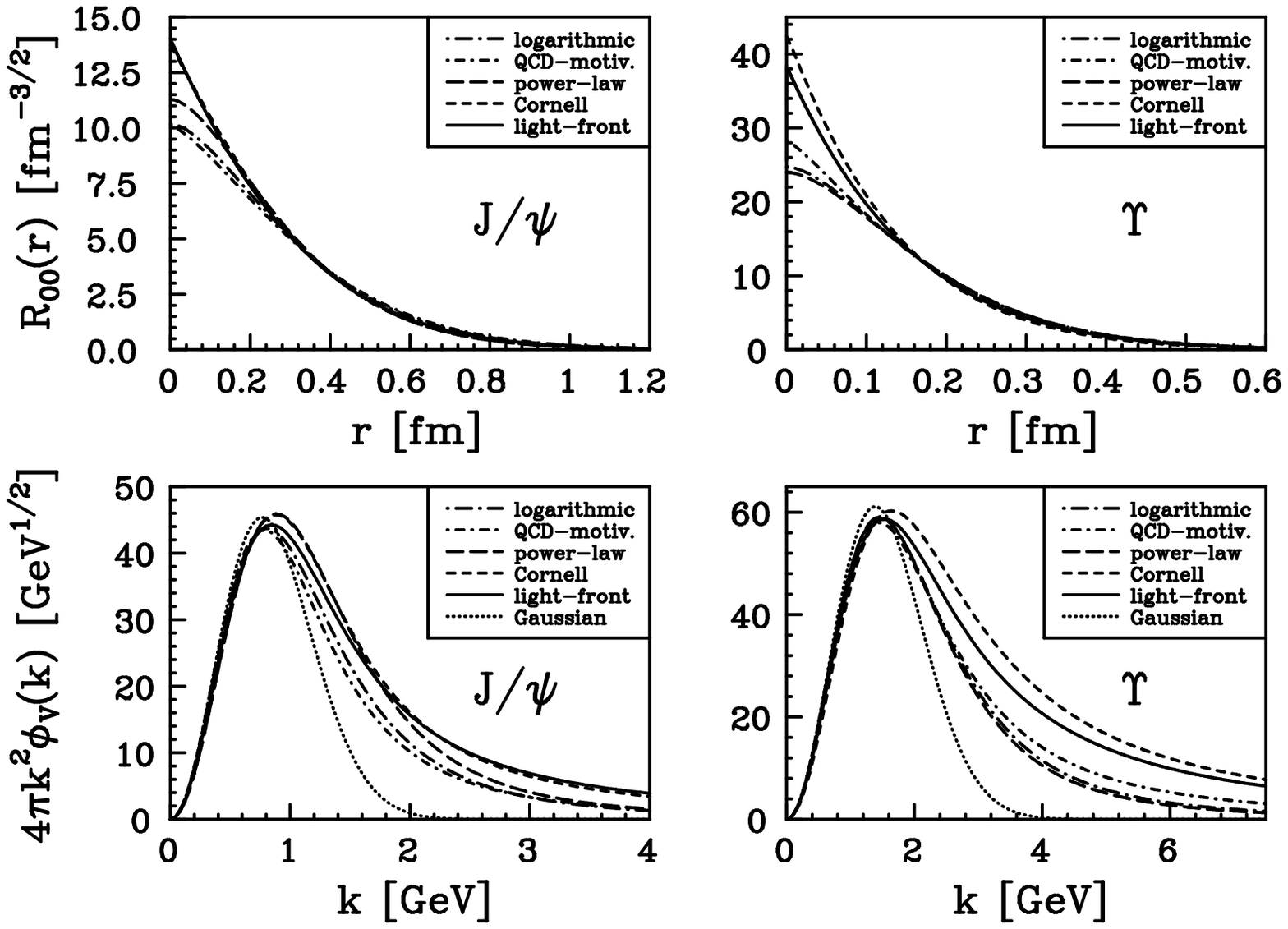}{6.0}{The non-relativistic quarkonium wave functions for the heavy
ground state mesons $J/\psi$ and $\Upsilon$ from various potential models
\protect\cite{charmlog,charmqcd,charmpow,charmcor} and a light-front QCD bound 
state calculation \protect\cite{perry}.  In the lower part of the
figure, we also show a Gaussian fit adjusted to reproduce $\phi_V(k)$ at small 
$k$ (dotted lines).}

Note that for our actual numerical calculations we will restrict our 
considerations to the models of Refs. \cite{charmlog}, \cite{charmqcd} and
\cite{perry} for which the mass 
of the constituent quark is close to the mass of the bare current quark, i.e., 
$m_c \approx 1.5$ GeV and $m_b \approx 5.0$ GeV.  This is necessary to keep a 
minimal correspondence with the QCD formulae for hard processes which are 
expressed through the distribution of bare quarks \cite{Brod94}.  

In Fig. 6, we show the contributions of the different regions in momentum space
to the integral $\int\!d^3k\,\phi_V(k)$ for the potential model of Ref. 
\cite{charmlog} (logarithmic potential).  This integral appears, for instance,
in the expression for the $V \rightarrow e^+e^-$ decay width.  Especially for 
$J/\psi$ mesons, the conventional non-relativistic potential models lead to a 
significant high momentum tail in the respective wave functions, and the
contribution of the relativistic region\footnote{Evidently, relativistic
effects should become important already at significantly smaller $k$.}
${v \over c} \geq 1$ (or $k \geq m$) to the integral $\int\!d^3k\,\phi_V(k)$ 
(the shaded area in Fig. 6) becomes large.  For the potential model of Ref.
\cite{charmlog}, the contribution of the relativistic region $k \geq m$ 
to the integral under consideration is 30\% for $J/\psi$ (and $\leq$ 10\% for 
$\Upsilon$).  Also, for the $J/\psi$, half of the integral comes from the 
region $k \geq 0.7\,m$.  This is in line with the QCD prediction of large 
relativistic corrections to the $c \bar c$ bound state equations 
\cite{buchmueller}, and it puts the feasibility of a non-relativistic 
description of heavy quarkonium production in high energy processes seriously 
into question.  

\figa{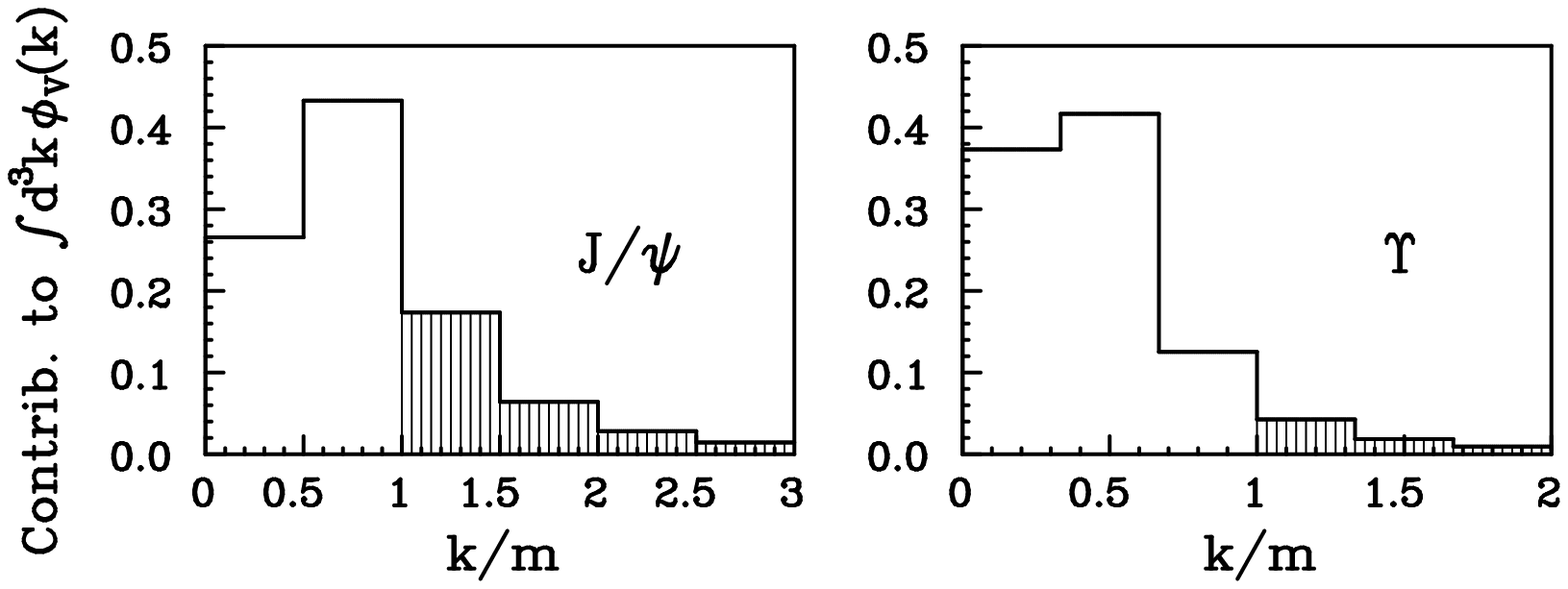}{6.0}
{Histogram of the relative contributions of the different 
regions in momentum space to the integral $\int\!d^3k\,\phi_V(k)$ for the 
potential model of Ref.  \protect\cite{charmlog}.}

The fact that, in particular for the $J/\psi$ meson, our numerical analysis 
yields a significant value for the high momentum component in the respective
non-relativistic wave functions is a very important result which should have
consequences far beyond the topic of diffractive vector meson production.
The large high momentum tail, visible in the lower part of Fig. 5, and the 
significant contribution of the relativistic region to the integral 
$\int\!d^3k\,\phi_V(k)$, displayed in Fig. 6, indicate that the $J/\psi$ meson
is not really a non-relativistic system!  This puts the non-relativistic
ansatz employed in the various potential models 
\cite{charmlog,charmqcd,charmpow,charmcor} as well as in the light-front QCD 
bound state calculation \protect\cite{perry} seriously into question.

However, there are more inconsistencies between the non-relativistic ansatz and
the hard reaction considered here.  For once, the requirement of 
self-consistency dictates that since in our formulae we use the gluon 
distribution $xG_N(x,Q^2)$ extracted from the data within a certain 
renormalization scheme ($\overline{MS}$), we are indebted to use the bare quark
mass defined within the same scheme.  This means that, in our final formulae in
Eqs. (\ref{eq0f}), (\ref{eq0g}), (\ref{eq0h}), (\ref{eq13}), (\ref{eq14}),
(\ref{eq14a}) and (\ref{eq14b}), the pole or constituent quark mass $m$ has to 
be replaced by the running mass $m_{run}(Q_{eff}^2)$, where \cite{running}
\begin{equation}
m^2~\rightarrow~
m_{run}^2(Q_{eff}^2)~=~m^2\left(1-{8\alpha_s\over 3\pi}\right) 
\ .
\label{eq15}
\end{equation}
Here, $\alpha_s$ is evaluated at $Q_{eff}^2$, i.e., at the 
effective scale of the reaction determined via the so-called ``rescaling of 
hard processes".  This is another consequence of the difference between soft, 
non-perturbative physics (as described, for instance, by non-relativistic 
quarkonium potential models) and hard perturbative QCD, and it further stresses
the inadequacies of a naive straightforward application of non-relativistic
potential models in this context.

Another mismatch between the soft non-relativistic and the hard light-cone 
approach appears within the evaluation of the $V \rightarrow e^+e^-$ decay 
width.  When $\Gamma_{V \rightarrow e^+e^-}$ is calculated from the 
non-relativistic wave function $\phi_V(k)$ via  
\begin{equation}
\Gamma_{V \rightarrow e^+e^-} =
{16\pi\alpha^2e_q^2\over M_V^2}\,
\left(1-{16\alpha_s\over 3\pi}\right)\,
\left|\,\int{d^3k\over (2\pi)^3}\,\phi_V(k)\,\right|^2
\ ,
\label{eq16}
\end{equation}
a QCD correction factor \cite{barbieri} appears, $1-{16\alpha_s\over 3\pi}$, 
which can be numerically large (${16\alpha_s\over 3\pi}\approx 0.35 -      
0.65$ for $J/\psi$), while no such term is present within the relation 
\cite{Brod94} with the light-cone $q\bar q$ wave function $\phi_V(z,k_t)$,
\begin{equation}
\Gamma_{V \rightarrow e^+e^-} = {32\pi\alpha^2e_q^2\over M_V}\,
\left|\,\int dz\,\int {d^2k_t\over 16\pi^3}\,\phi_V(z,k_t)\,\right|^2
\ .
\label{eq17}
\end{equation}
The appearance of this correction factor is the main differences between the 
various non-relativistic potential models and a ``true" QCD approach, in which
the light-cone wave function of the minimal $q \bar q$ Fock component in the 
vector meson is employed.  It is a radiative correction to the matrix 
element of the electromagnetic current calculated, essentially, while neglecting 
quark Fermi motion effects.  The respective Feynman diagram is shown in Fig. 7. 

\figa{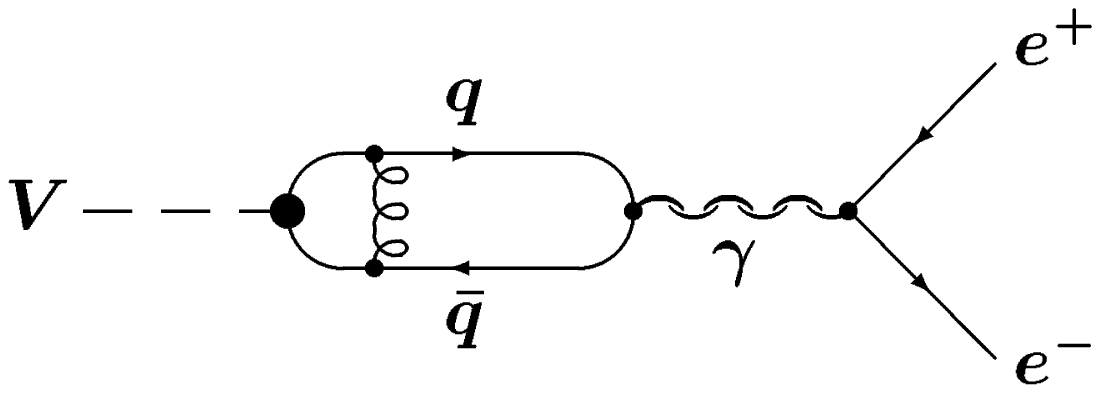}{3.5}
{The QCD radiative correction \protect\cite{barbieri} to the
$V \rightarrow e^+e^-$ decay width.}

The correction arises from the exchange of a gluon between the quark and the
antiquark in the vector meson with fairly large transverse momentum, $\langle 
l_t \rangle \approx m$.  The physical interpretation of the $1-{16\alpha_s\over
3\pi}$ correction factor is that it ``undresses" the constituent quarks, which 
are the relevant degrees of freedom of the non-relativistic wave function, back
to current quarks, which, in turn, are the degrees of freedom the light-cone 
wave function refers to and to which the $V \rightarrow e^+e^-$ decay width is
ultimately connected.  This, once more, underlines the limits of applicability 
of non-relativistic potential models in that context.  Note that this radiative
correction is also present in the light-front QCD bound state calculation 
of Ref. \cite{perry} because also there the relevant degrees of freedom are 
dressed constituent and not bare current quarks.

\subsection{The light-cone wave function}
\label{III.B}

Leaving these issues behind for the moment,  we can, in principle, deduce a
light-cone wave function $\phi_V(z,k_t)$ appropriate for the evaluation of
time-ordered perturbation theory diagrams from the non-relativistic wave 
function $\phi_V(k)$.  This requires a translation of conventional 
non-relativistic diagrams into light-cone perturbation theory diagrams.  This, 
in turn, can be achieved by the purely kinematical identification of the 
Sudakov variable $z$, which denotes the fraction of the plus-component of the 
meson's momentum carried by one of the quarks, with
\begin{equation}
z~=~{1\over 2}\left(1+{k_z\over\sqrt{k^2+m^2}}\right) \ .
\label{eq18}
\end{equation}
This yields
\begin{eqnarray}
k^2~&\longrightarrow&~{k_t^2+(2z-1)^2m^2 \over 4z(1-z)} \ ,
\label{eq19}
\\
d^3k~&\longrightarrow&~{\sqrt{k_t^2+m^2}\over
4[z(1-z)]^{3/2}}\ dz\,d^2k_t \ ,
\label{eq20}
\end{eqnarray}
where $\pm \vec k_t$ are the quarks' transverse momenta.  This, together with 
the conservation of the overall normalization of the wave function,
\begin{equation}
1~=~\int {d^3k\over (2\pi)^3}\,|\phi_V(k)|^2
 ~=~\int dz \int {d^2k_t \over 16\pi^3}\,|\phi_V(z,k_t)|^2
\label{eq21}
\end{equation}
then gives a relationship between the light-cone and the non-relativistic wave 
function:
\begin{equation}
\phi_V(z,k_t)~=~^4\!\sqrt{{k_t^2+m^2 \over 4 [z(1-z)]^3}}~
                  \phi_V\!\left(k=\sqrt{{k_t^2+(2z-1)^2m^2
                                  \over 4z(1-z)}}         \right)
\ .
\label{eq22}
\end{equation}
From $\phi_V(z,k_t)$ we then calculate the quarkonium's wave function in 
transverse impact parameter space $\phi_V(z,b)$ through a two-dimensional
Fourier transformation,
\begin{equation}
\phi_V(z,b)~=~
      \int {d^2k_t\over 16\pi^3}~e^{i{\vec k_t}\cdot{\vec b}}~\phi_V(z,k_t)
\ .
\label{eq23}
\end{equation}

Obviously, the non-relativistic quarkonium model, designed as a description of
the $q\bar q$ constituent quark component -- including the $1-{16\alpha_s\over
3\pi}$ factor which accounts for radiative corrections -- does not include 
gluon emission at a higher resolution.  So it is not surprising that the 
$\phi_V(z,b)$ that we find does not display the expected asymptotic behavior 
\cite{CZ},
\begin{equation}
\phi_V(z,b=0)~\not\propto~z(1-z) \ ,
\label{eq24}
\end{equation}
This is illustrated in Fig. 8.  There, we compare the quarkonium wave functions
$\phi_V(z,b=0)$ obtained in that manner from the non-relativistic potential 
models of Refs. \cite{charmlog}, \cite{charmqcd} and \cite{perry} 
with a hard wave function 
$\phi_V^{hard}(z,b=0)=a_0 z(1-z)$, where the parameter $a_0$ was adjusted by 
means of Eq. (\ref{eq17}) to reproduce the vector meson's leptonic decay width 
$\Gamma_{V\rightarrow e^+e^-}$.
For transversely polarized vector mesons, the light-cone wave function
should behave as $\propto z^2(1-z)^2$.  This  follows from the
analysis of respective pQCD diagrams.
  
\figa{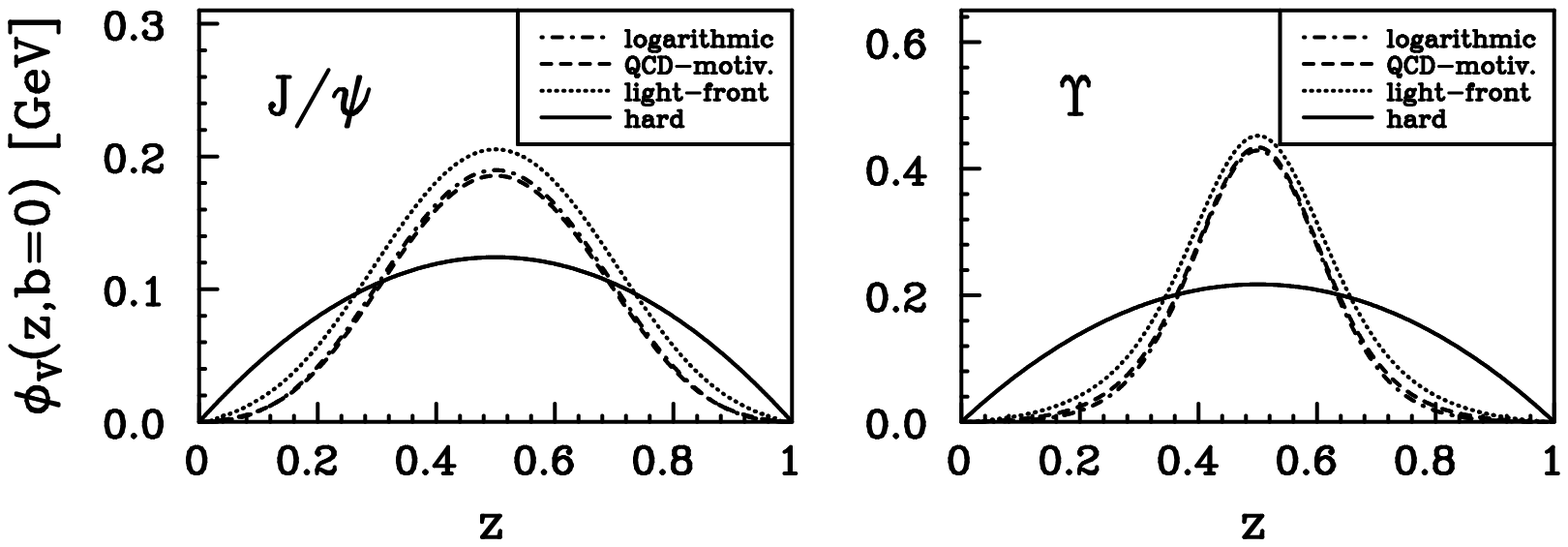}{6.5}{The quarkonium wave functions, $\phi_{J/\psi}(z,b)$ and
$\phi_{\Upsilon}(z,b)$, for $b=0$.  The dot-dashed, dashed and dotted lines 
correspond to the non-relativistic potential models of Refs.  
\protect\cite{charmlog}, \protect\cite{charmqcd} and \protect\cite{perry}, 
respectively, and the solid lines refer to the ``hard physics" limit 
$\phi_V(z,b=0)\propto z(1-z)$.}

\subsection{Hard physics}
\label{III.C}

We argued in subsection \ref{III.A} 
that, although the average properties of heavy 
quarkonium bound states might, in general, be quite well described within 
a non-relativistic framework due to the large value of the quark mass, 
there appear significant high momentum (or ``hard physics") corrections
if observables are considered which crucially depend on short distances.

One example is the leptonic decay width, $\Gamma_{V \rightarrow e^+e^-}$,
which acquires large radiative corrections in a non-relativistic potential 
model.  An analysis of the respective Feynman diagram, shown in Fig. 7, yields 
that these corrections arise from relativistic momenta, $k \gtrsim m$.  Even 
putting those corrections aside, already the quantity which is related to the 
decay width in zeroth order, $\int\!d^3k\,\phi_V(k)$, contains large 
contributions from the relativistic region $k \geq m$ (30\% for $J/\psi$ for 
a typical potential model).  And when we (purely kinematicly) translate the 
non-relativistic wave functions into light-cone coordinates, we find that they 
do not display the expected asymptotic short distance behavior $\phi_V(z,b=0) 
\propto z(1-z)$ as dictated by perturbative one-gluon exchange.

This suggests that the non-relativistic potential model wave functions might 
describe the $q \bar q$ leading Fock state in heavy quarkonia for fairly large 
(average) distances, but the description breaks down in the limit of small 
distances or high momenta.  As these play a crucial role for the processes we 
are interested in, we designed the following strategy:  

First, we extract a light-cone wave function from a non-relativistic potential 
model through the purely kinematical transformations of Eqs. (\ref{eq18}) 
through (\ref{eq22}), which we then Fourier transform into transverse impact 
parameter space via Eq. (\ref{eq23}).  However, we have confidence in that wave
function, which we denote $\phi_V^{NR}(z,b)$, only for transverse distances 
$b \gtrsim {1 \over m}$, and we expect it to be modified at shorter distances 
by means of the ``hard physics" corrections discussed in the above.  We thus 
set
\begin{equation}
\smallskip
\phi_V(z,b)~=~\cases{\phi_V^{NR}(z,b)         &for $b \geq b_0$\ , \cr
                     \phi_V^{LC}(z,b)         &for $b <    b_0$\ , \cr}
\label{eq25}
\smallskip
\end{equation}
where $b_0 \sim {1\over m}$.

The wave function $\phi_V^{LC}(z,b)$ is then constructed such that: 1) 
$\phi_V(z,b)$ and $\partial\phi_V(z,b)/\partial b$ are continuous at $b=b_0$,
2) $\phi_V^{LC}(z,b)$ has the correct asymptotic behavior dictated by the 
perturbative exchange of hard gluons, i.e., $\phi_V^{LC}(z,b=0)\propto z(1-z)$,
and 3) $\phi_V^{LC}(z,b)$ reproduces the vector meson's leptonic decay width
{\em without} account of the radiative correction term $1-{16\alpha_s\over
3\pi}$, i.e., Eq. (\ref{eq17}) is used to calculate $\Gamma_{V \rightarrow 
e^+e^-}$.  We expand $\phi_V^{LC}(z,b)$ in terms of Gegenbauer 
polynomials\footnote{The expansion in Gegenbauer polynomials has
 nothing to do
with renormalization group methods.  They provide a complete basis for the 
$\phi_V(z,b)$ under consideration and allow a smooth interpolation between the
$b \to 0$ and $b \gtrsim 1/m$ regimes.  The series in Eq. (\ref{eq26}) is
terminated when convergence is achieved, which, in practice, is at
$n \sim 10$.},
\begin{equation}
\phi_V^{LC}(z,b)~=~a_0(b)\,z(1-z)\,\left(1~+~
                      {\displaystyle \sum_{n=2,4,\ldots}}\!a_n(b)\,
                      C^{3/2}_n(2z-1) \right)
\ ,
\label{eq26}
\end{equation}
and we assume that the coefficients $a_i(b)$ depend on $b^2$ only through
second order, i.e., $a_i(b) = a_{i0} + a_{i1} b^2 + a_{i2} b^4$.  This, 
together with the conditions 1) through 3) is sufficient to unambiguously 
determine $\phi_V^{LC}(z,b)$.  In our actual numerical calculations, we set
$b_0=0.3$ fm for $J/\psi$ and $b_0=0.1$ fm for $\Upsilon$.  Respective
wave functions are shown in Fig. 9.  The dot-dashed, dashed and dotted lines
show the non-relativistic wave functions $\phi_V^{NR}(z={1\over 2},b)$ before 
the ``hard physics" corrections discussed in this subsection were imposed, and 
the solid lines depict the modified wave functions 
$\phi^{LC}_V(z={1\over 2},b)$ of Eq. (\ref{eq26}).

\figa{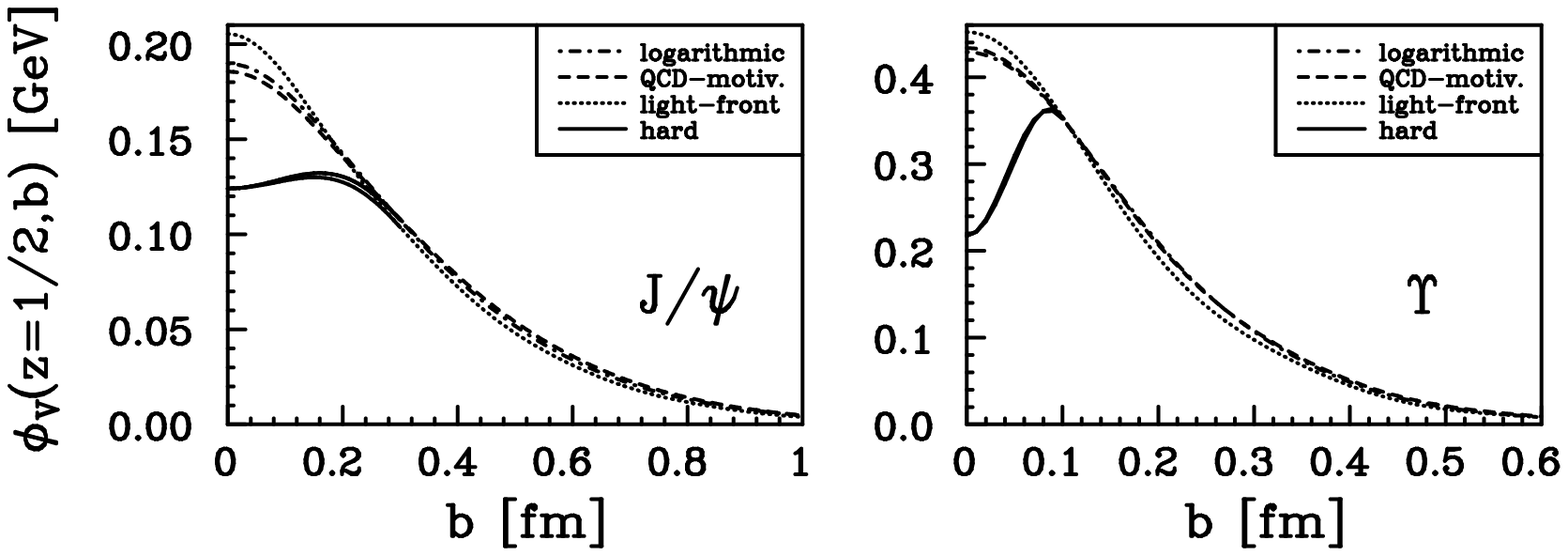}{6.5}
{The quarkonium wave functions, $\phi_{J/\psi}(z,b)$ and
$\phi_{\Upsilon}(z,b)$, for $z={1\over 2}$.  The dot-dashed, dashed and dotted 
lines correspond to the non-relativistic potential models of Refs.  
\protect\cite{charmlog}, \protect\cite{charmqcd} and \protect\cite{perry}, 
and the solid lines refer to the inclusion of the ``hard physics" 
corrections of Eqs. (\protect\ref{eq25}) and (\protect\ref{eq26}) for
$b<b_0$.  We set $b_0=0.3$ {\rm fm} for $J/\psi$ and $b_0=0.1$ {\rm fm} for 
$\Upsilon$.}

Note that the ``hard physics" corrections, which we introduced in the above,
address effects that are of higher order in an expansion in ${1\over m}$.  But
the prescription of modifying the wave function at $b < b_0$ only accounts
for some (but not all) corrections to this order.  We thus emphasize that
the corrections outlined in this subsection are a model estimate only.

\subsection{Vector meson production}
\label{III.D}

For the potential models of Refs. \cite{charmlog}, \cite{charmqcd} and
\cite{perry}, the 
non-relativistic wave functions $\phi_V^{NR}(z,b)$ yield values for the 
asymptotic correction factor $\eta_V$ of Eq. (\ref{eq0b}) of $\eta_{J/\psi} 
\approx 2.3 - 2.4$ and $\eta_{\Upsilon} \approx 2.1 - 2.2$ if the 
``hard physics" correction outlined in the last section is not considered.  
While, with that correction, they yield $\eta_V=3$.  Note that the static
limit, i.e., $\phi_V(z,b) = \delta(z-1/2)\phi_V(b)$, gives $\eta_V=2$.  

Furthermore, in line with the discussion in subsection \ref{III.A}, 
we do not use the
pole mass $m$ in our final formulas, but we replace it with the running mass 
$m_{run}$, as given by Eq. (\ref{eq15}).  We can then use the wave functions
$\phi_V(z,b)$, that we constructed in the last two subsections, to calculate 
the correction factors of Eqs. (\ref{eq0g}) and (\ref{eq0h}) or (\ref{eq13}) 
and (\ref{eq14}).  Putting 
everything together, we can rewrite the forward differential cross section for 
photo- and electroproduction of heavy vector mesons of Eq. (\ref{eq0f}) as the 
product of an asymptotic expression and a finite $Q^2$ correction, 
${\cal C}(Q^2)$, where
\begin{eqnarray}
&&\left. {d\sigma_{\gamma^{(*)}N\rightarrow VN}\over dt}\right|_{t=0}~=~
\nonumber
\\
&&\quad=~{12\pi^3 \Gamma_V M_V^3\over \alpha_{EM}(Q^2+4m^2)^4}
  \,\left|\alpha_s(Q_{eff}^2)\left(1+i\beta\right)xG_N(x,Q_{eff}^2)\right|^2
  \,\left(1+\epsilon {Q^2\over M_V^2}\right)
  \,{\cal C}(Q^2)
\ ,
\label{eqx1}
\end{eqnarray}
with 
\begin{equation}
{\cal C}(Q^2)~=~\left({\eta_V\over 3}\right)^{\!2}
              \,\left({Q^2+4m^2 \over Q^2+4m_{run}^2}\right)^{\!\!4}
              \,T(Q^2)
               ~{R(Q^2)+\epsilon {Q^2\over M_V^2}
                \over 1+\epsilon {Q^2\over M_V^2}}
\ .
\label{eqx2}
\end{equation}
Here, $\eta_V$ is the leading twist correction of Eq. (\ref{eq0b}),
the factor $T(Q^2)$ of Eq. (\ref{eq0g})
accounts for effects related to the quark motion in the produced vector meson, 
$\epsilon$ is the (virtual) photon's polarization, and the factor $R(Q^2)$ 
of Eq. (\ref{eq0h}) parameterizes the relative contribution of the transverse 
polarization.  The pole mass $m$ we set to $m=1.5$ GeV for $J/\psi$ and to 
$m=5.0$ GeV for $\Upsilon$ production, and $m_{run}$ is the ``running mass"
of Eq. (\ref{eq15}) which, through $Q^2_{eff}$, depends on $Q^2$ and the 
vector meson's wave function.

Results for the Fermi motion suppression factor, $T(Q^2)$ of Eq. (\ref{eq0g}),
and the finite $Q^2$ correction, ${\cal C}(Q^2)$ of Eq. (\ref{eqx2}), are shown
in Fig. 10 for $J/\psi$ and $\Upsilon$ photo- and electroproduction.
The calculations
are based on vector meson wave functions from the models of Refs. 
\cite{charmlog} \cite{charmqcd} and \cite{perry}.  
The solid line, labeled hard, refers to the inclusion of the ``hard physics" 
corrections of Sect. \ref{III.C}.  For the evaluation of ${\cal C}(Q^2)$, the 
photon's polarization $\epsilon$ was set to $1$.

\figa{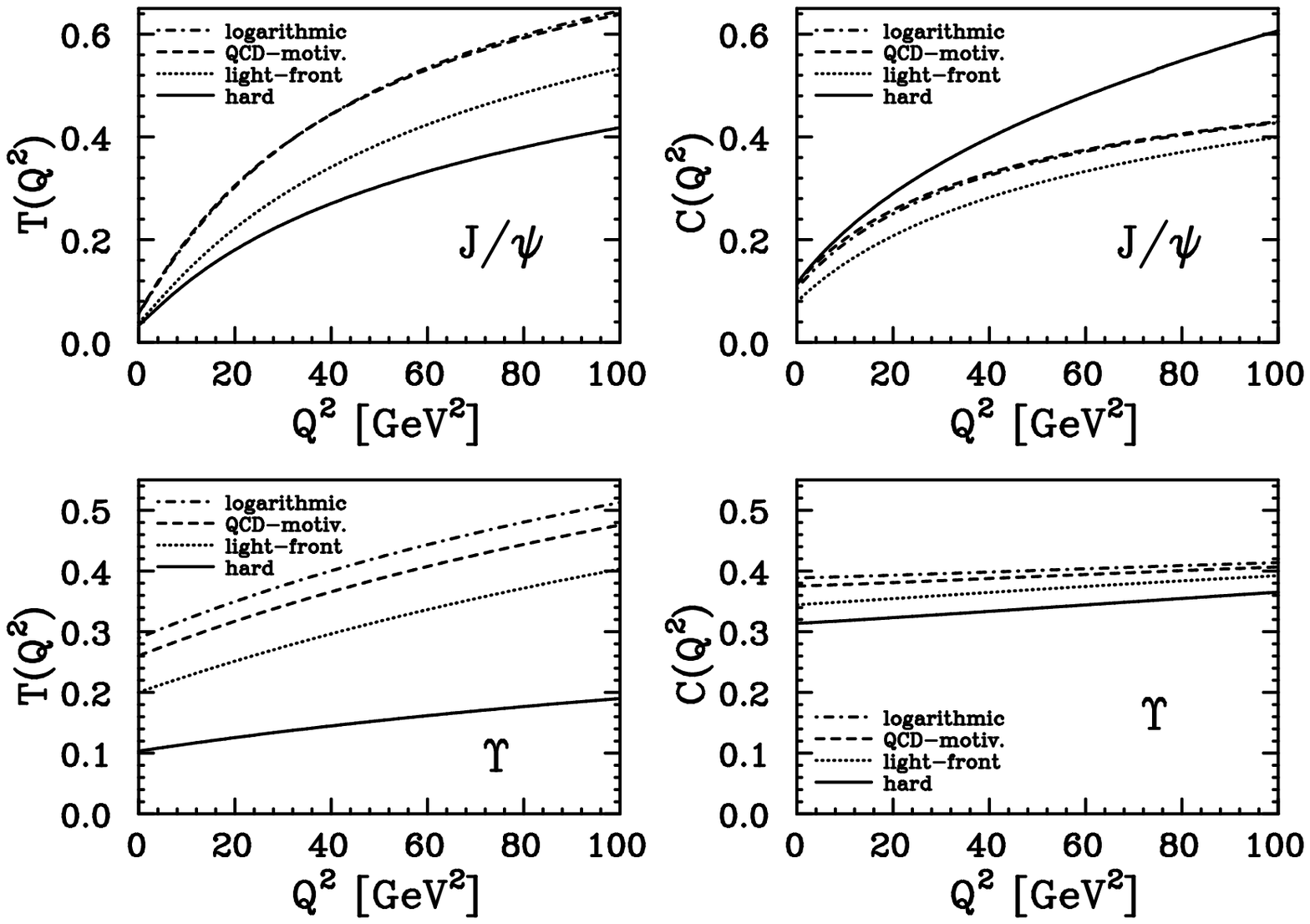}{6.5}
{The Fermi motion suppression factor, $T(Q^2)$ of Eq. 
(\protect\ref{eq0g}) and the finite $Q^2$ correction, ${\cal C}(Q^2)$ of Eq. 
(\protect\ref{eqx2}), for $J/\psi$ and $\Upsilon$ production.}

It can be seen from that figure that, for reasonable $Q^2$, the correction 
factor ${\cal C}(Q^2)$, which measures the suppression of the cross section 
due to the quark motion in the produced vector meson, is significantly smaller 
than $1$.  This shows that the asymptotic expression, i.e., Eq. (\ref{eqx1}) 
with the finite $Q^2$ correction ${\cal C}(Q^2)$ set to $1$, is valid for 
extremely large $Q^2$ only.  Note that the ``hard physics" corrections of Sect.
\ref{III.C}
lead to a stronger suppression in $T(Q^2)$, but, at least for $J/\psi$
production, to less suppression in the final correction factor ${\cal C}(Q^2)$.
The reason for this is, firstly, that the ``hard physics" correction increases 
$\eta_V$ of Eq. (\ref{eq0b}) from around $2.1 - 2.4$ to 3, and, secondly,
that the running mass $m_{run}$ of Eq. (\ref{eq15}) is smaller than the pole 
mass, which also enhances the cross section.  In addition, the relative 
contribution of the transverse polarizations $R(Q^2)$ of Eq. (\ref{eq0h}) 
is very close to $1$ both for $J/\psi$ and $\Upsilon$ production for all 
experimentally accessible $Q^2$ if the ``hard physics" corrections are left
out.  However, at least for $J/\psi$ production, after the ``hard physics" 
corrections are considered, $R(Q^2)$ increases significantly with $Q^2$.
This, together with the changes  through $\eta_V$ and $m_{run}$ lead to
the difference between $T(Q^2)$ and ${\cal C}(Q^2)$.  The cross sections
are enhanced also due to the so-called ``rescaling of hard processes", 
because the virtuality that enters in the gluon density, $Q^2_{eff}$ of Eq. 
(\ref{eq8}), is larger than the naive estimate of ${\overline Q}^2={Q^2+M_V^2
\over 4}$.   This was discussed in detail in Section \ref{II.C}.

Note that for photoproduction, i.e., for $Q^2$=0, only the transverse 
polarizations
are present, and the correction ${\cal C}(Q^2)$ of Eq. (\ref{eqx2}) takes on 
the form
\begin{equation}
{\cal C}(0)~\propto~T(0)\,R(0)~\propto~
\left[{\int\!{dz\over z^2(1-z)^2}
        \int\!d^2k_t~\phi_V(z,k_t)~\Delta_t~\phi_\gamma(z,k_t)
       \over
       \int\!{dz\over z(1-z)}\int\!d^2k_t~\phi_V(z,k_t)}                  
 \right]^2
\ .
\label{eqx3}
\end{equation}
The presence of the 
${1\over z^2(1-z)^2}$
term strongly enhances smearing in the longitudinal motion, i.e., 
the contribution of asymmetric $q\bar q$ pairs with 
$z \ne {1 \over 2}$ is pronounced.

\subsection{The ratio of $\Upsilon$ and $J/\psi$ photoproduction}
\label{III.D1} 

One can furthermore conclude from Fig. 10, together with our master formula in 
Eqs. (\ref{eqx1}) and (\ref{eqx2}), that, after an eventual luminosity upgrade,
a significant production of $\Upsilon$ mesons is expected at HERA.  The cross 
section ratio of $\Upsilon$ to $J/\psi$ photoproduction (at fixed
 $x$) is 
approximately 
\begin{equation}
{\sigma(\gamma+p\rightarrow ~\Upsilon~ +p)\over 
\sigma(\gamma+p\rightarrow~J/\psi~+p)}~\approx~
{\Gamma_{\Upsilon}~M_{\Upsilon}^3~m_c^8 \over \Gamma_{J/\psi}M_{J/\psi}^3m_b^8}
\cdot
{|\alpha_s(1+i\beta)xG_N(Q_{eff}^2[\Upsilon])|^2 \over
 |\alpha_s(1+i\beta)xG_N(Q_{eff}^2[J/\psi])|^2}
\cdot
{{\cal C}_\Upsilon(0) \over {\cal C}_{J/\psi}(0)}
\ .
\label{eq27}
\end{equation}
The first factor on the right hand side of Eq. (\ref{eq27}) is the
dimensional
estimate, and it yields a relative suppression of $\Upsilon$ photoproduction
of about $1:2000$ if we set for the quark masses $m_c=1.5$ GeV and $m_b=5.0$
GeV.  The second term arises due to the so-called ``rescaling of hard 
processes", and it enhances the 
cross section ratio by a factor of about $3$ for $x=10^{-3}$. 
The third term is connected to the wave function dependent effects, and it 
enhances the production ratio also by a factor of about $3$.  All together, the
cross section for $\Upsilon$ photoproduction is suppressed by approximately 
$1:200$ as compared to $J/\psi$ photoproduction for the same $x$.
For the same $W$, an extra suppression factor
$\approx (M_{\Upsilon}/M_{J/\psi})^{0.8} \approx 2.4$ is present.
Note that the different $Q^2$ scale 
and higher twist effects (the ``rescaling of hard processes" 
as well as the 
${\cal C}(Q^2)$ correction) increase the relative yield that we predict by 
about an order of magnitude as compared to the naive dimensional estimate!

\section{The $J/\Psi$ photo- and electroproduction cross section}
\label{IV}

In Fig. 11, we compare our predictions\footnote{A respective FORTRAN program is
available by request from koepf@mps.ohio-state.edu or via the WWW at
http://www.physics.ohio-state.edu/\~{}koepf.} for the $J/\Psi$ photoproduction 
cross section with the data.  We used a slope parameter of $B_{J/\Psi}
= 3.8$ GeV$^{-2}$, as measured by the H1 collaboration \cite{H1},
to calculate the total cross section from our predictions for the forward
differential cross section at $t=0$, and the Fermi motion corrections
and the ``rescaling of hard processes" are accounted for.  For the former, the 
charmonium potential of Ref. \cite{charmlog} was employed and the ``hard 
physics" corrections, as outlined in Sect. \ref{III.C}, were taken 
into account.
We furthermore replaced the quark pole mass with the running mass $m_{run}$ 
from Eq. (\ref{eq15}), and we set $x={Q^2+M_V^2\over W^2}$.  The formulas
to obtain the forward differential cross section are given in Eqs. (\ref{eqx1})
and (\ref{eqx2}).

\figa{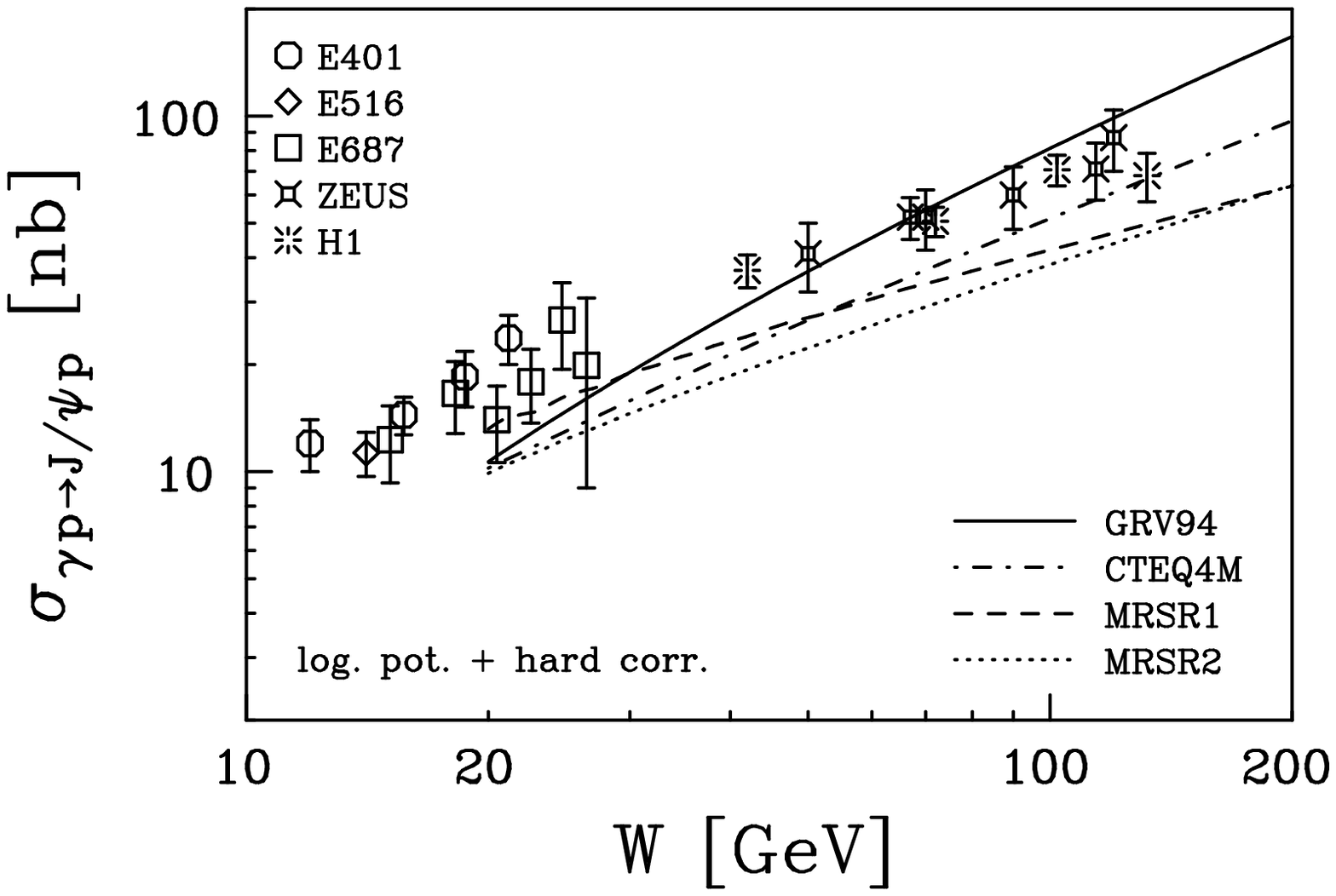}{4.0}
{The $J/\Psi$ photoproduction cross section for several recent 
parameterizations of the gluon density \protect\cite{GRV,MRS,CTEQ} in 
comparison with experimental data from E401 \protect\cite{e401}, 
E516 \protect\cite{e516}, E687 \protect\cite{e687}, ZEUS '93 
\protect\cite{ZEUS1}, and H1 \protect\cite{H1phot}.}

As can be seen from Fig. 11, the predictions of our pQCD calculation agree
with the data within the uncertainties in the nucleon's gluon density,
and the energy dependence of the data is much better reproduced within the
pQCD picture, where $\sigma \propto W^{0.7 - 0.8}$, than through the
soft Pomeron model \cite{pom}, where $\sigma \propto W^{0.32}$.  A rough fit 
\cite{H1} to the data depicted in Fig. 11 yields $\sigma \propto W^{0.9}$.

To investigate the $Q^2$ dependence of $J/\Psi$ production, we show in Fig.
12 the ratio of the electro- to photoproduction cross sections, i.e., we plot
$\sigma_{\gamma^*+p \rightarrow J/\psi+p}(Q^2)
\over\sigma_{\gamma+p \rightarrow J/\psi+p}(Q^2=0)$ 
as a function of the virtuality of the photon.  In 
particular, we compare a calculation where the Fermi motion corrections were
left out (dotted lines), with an evaluation were the latter effects were 
included while either using just the non-relativistic wave functions (dashed 
lines) or also accounting for the ``hard physics" corrections (solid lines).  

\figa{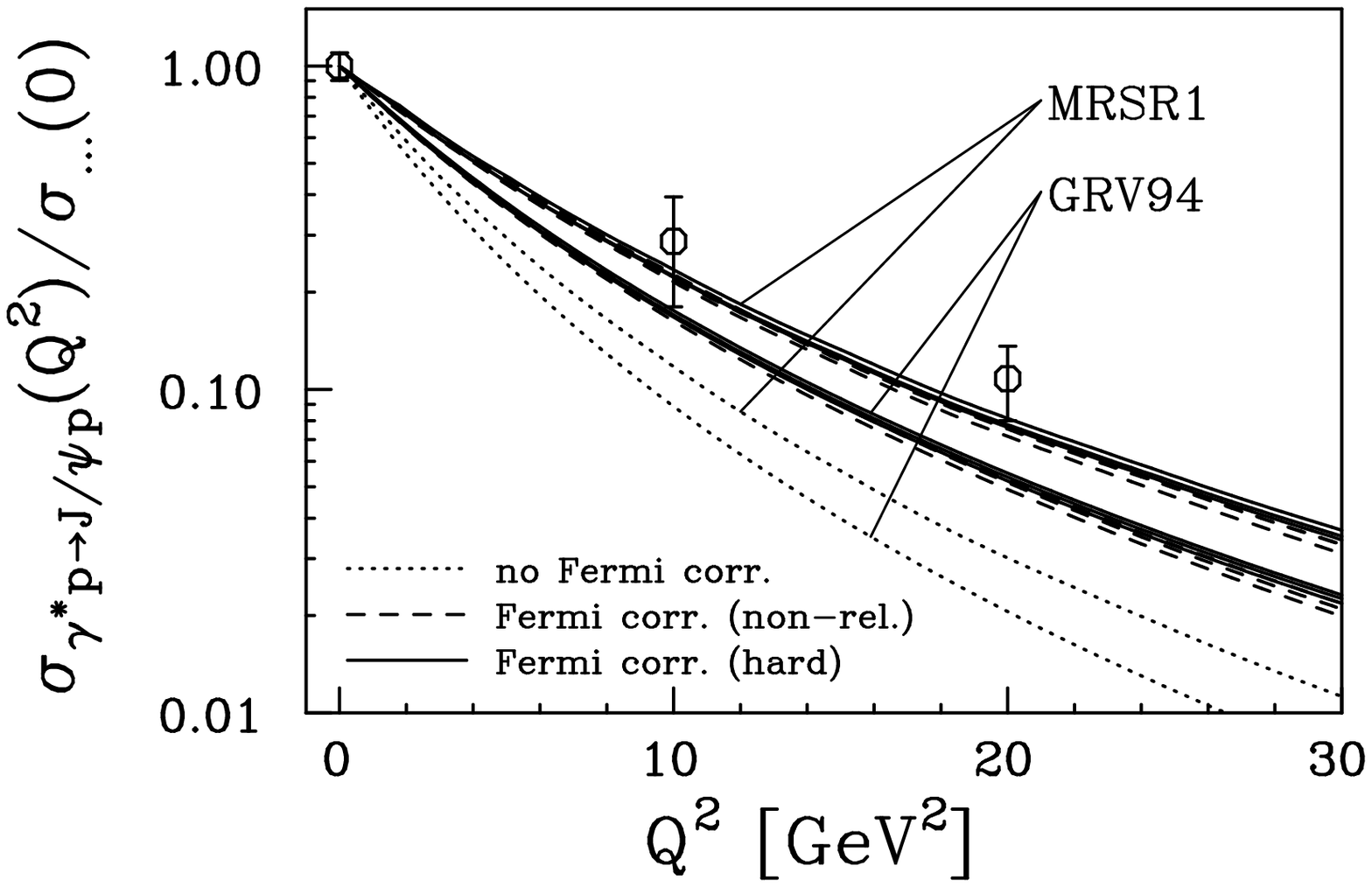}{4.0} 
{The ratio of the $J/\Psi$ electro- to photoproduction cross
section for two recent parameterizations of the gluon density 
\protect\cite{GRV,MRS} and for various potential models 
\protect\cite{charmlog,charmqcd,perry} in comparison with experimental data 
from H1 \protect\cite{H1phot}.}

As can be seen from Fig. 12, the Fermi corrections are necessary to achieve
agreement with the data.  However, at this point, the quality of the data
is not sufficient to distinguish between the various potential models or to
decide whether the ``hard physics" corrections which were imposed on those
wave functions at small transverse inter-quark distances -- see Sect. 
\ref{III.C} --
lead to an improvement.  This should change if the '95 data, which have much 
better statistics,  become available.  The fact that we somewhat underestimate 
the $J/\psi$ photoproduction cross section -- see Fig. 11 -- and, at the same 
time, overestimate the $Q^2$ dependence of $J/\psi$ electroproduction -- see
Fig. 12 -- suggests that our quark motion correction factor ${\cal C}(Q^2)$ of 
Eq. (\ref{eqx2}) is too small at $Q^2=0$ and it falls off 
too quickly at larger $Q^2$.  This implies that the wave functions 
which we use 
fall off too slowly in transverse momentum space and they are too steep as a 
function of the impact parameter $b$, i.e., the respective $\langle k^2_t 
\rangle$ is too large.

\section{The $\rho^o$ electroproduction cross section}
\label{V}

Although the main topic of this work is heavy meson photo- and 
electroproduction, we still consider an update of our predictions of Ref. 
\cite{hepph} in regards to $\rho^o$ electroproduction warranted in light 
of the new data as well as theoretical developments in that realm.
Currently, absolute cross sections for exclusive $\rho$-meson production
are available from NMC \cite{NMC}, ZEUS \cite{ZEUS}, and H1 \cite{H1},
and preliminary results exist from ZEUS from the 1994 run \cite{ZEUS94}.
From our predictions for the forward differential cross section, 
$\left.{d\sigma_{\gamma_L^* p \rightarrow \rho p}\over dt}\right|_{t=0}$
of Eq. (\ref{eq0c}),  the total cross section was calculated using a slope 
parameter of $B_\rho=5$ GeV$^{-2}$.  This is consistent with the values given 
by the NMC \cite{NMC} ($4.6 \pm 0.8$ GeV$^{-2}$) and ZEUS \cite{ZEUS} ($5.1 
\pm 1.2$ GeV$^{-2}$) collaborations and slightly smaller than that obtained 
by H1 \cite{H1} ($7.0 \pm 0.8$ GeV$^{-2}$).

To indicate separately the spread 
that arises from the different available gluon densities and the uncertainty 
that stems from the various proposed $\rho$-meson wave functions, the 
theoretical predictions are shown for two (extremal) gluon densities,
GRV94(HO) of Ref. \cite{GRV} and MRSR2 of Ref. \cite{MRS}, and two different
wave functions, termed ``soft" and ``hard".  The ``soft" wave function refers
to a $\phi_\rho(z,k_t) \propto \exp\!\left(-{A k_t^2\over z(1-z)}
\right)$ with an average transverse quark momentum of
$\langle k^2_t \rangle = 0.18$ GeV$^2$ as extracted from a
QCD sum rule analysis by Halperin and Zhitnitsky \cite{halperin}, and the 
``hard" wave function corresponds to a $\phi_\rho(z,k_t) \propto z(1-z)\, {A 
\over (k_t^2+\mu^2)^2}$ obtained in another QCD sum rule analysis (for pions)
by Lee, Hatsuda and Miller \cite{miller}.  For the latter, $\langle k^2_t 
\rangle = 0.09$ GeV$^2$.  As outlined in detail in the above,
the wave function enters through the Fermi motion suppression factor,
$T(Q^2)$ of Eq. (\ref{eq0d}).  $T(Q^2)$ is depicted in Fig. 13 for various
available $\rho$-meson wave functions: ``hard" and ``soft" were discussed in
the above, ``soft$_1$" refers to a duality wave function of the form
$\Theta\!\left(s_0-{k_t^2\over z(1-z)}\right)$ with $\langle k^2_t \rangle = 
0.15$ GeV$^2$ obtained in Ref. \cite{radyu2} 
and used for a similar analysis in \cite{rad},
and ``soft$_2$" labels a two-peak
Gaussian favored in the analysis of Halperin and Zhitnitsky \cite{halperin}.  
Note that the 
latter wave function seems quite extreme as it would correspond to 
a transverse spread of the $q\bar q$ component which is
larger than the meson's size!

\figa{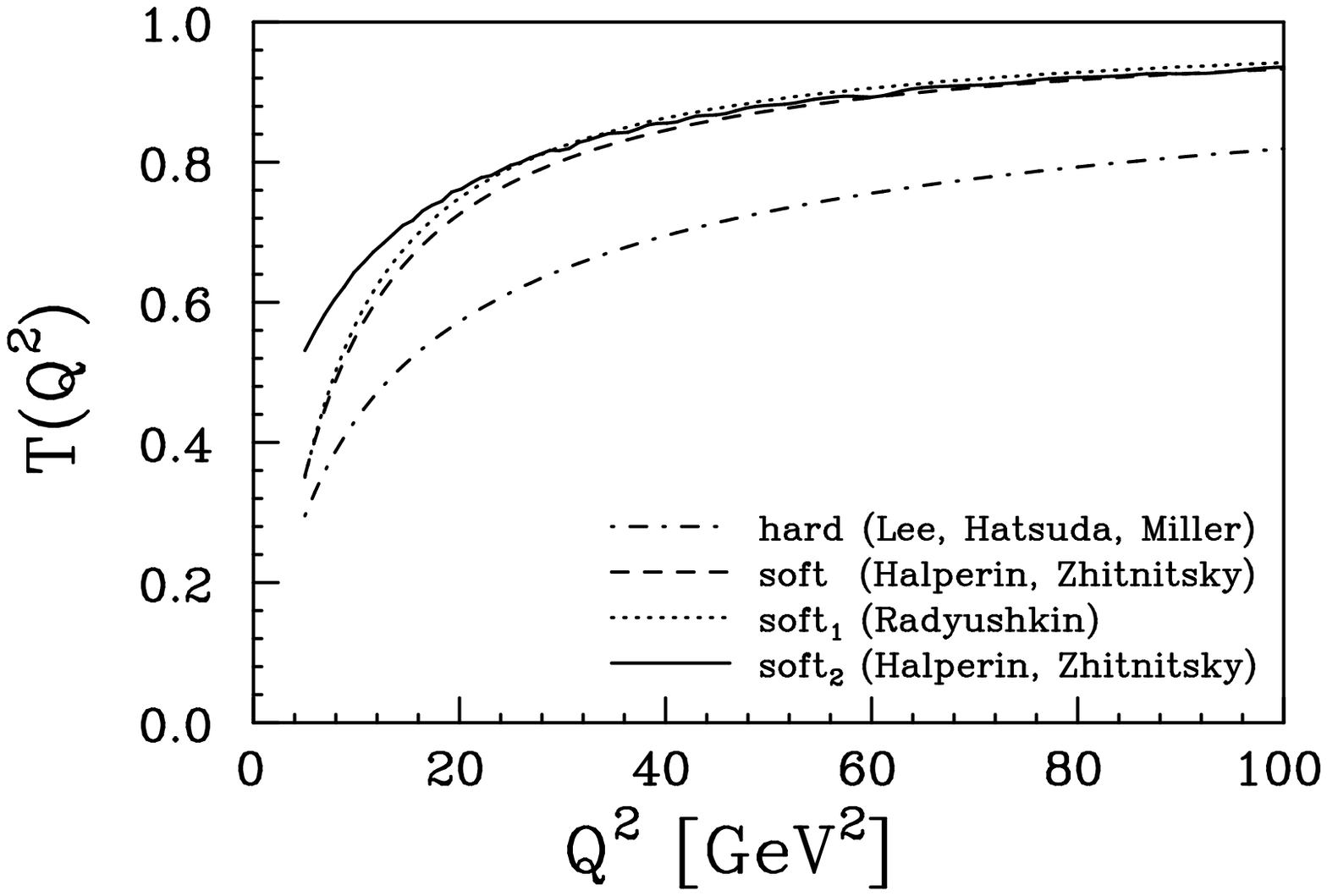}{4.0}
{The Fermi motion suppression factor $T(Q^2)$ of Eq.
(\protect\ref{eq0d}) for $\rho^o$ electroproduction for various $\rho$-meson
wave functions from Refs. \protect\cite{miller}, \protect\cite{halperin} and
\protect\cite{radyu2}.}

\figp{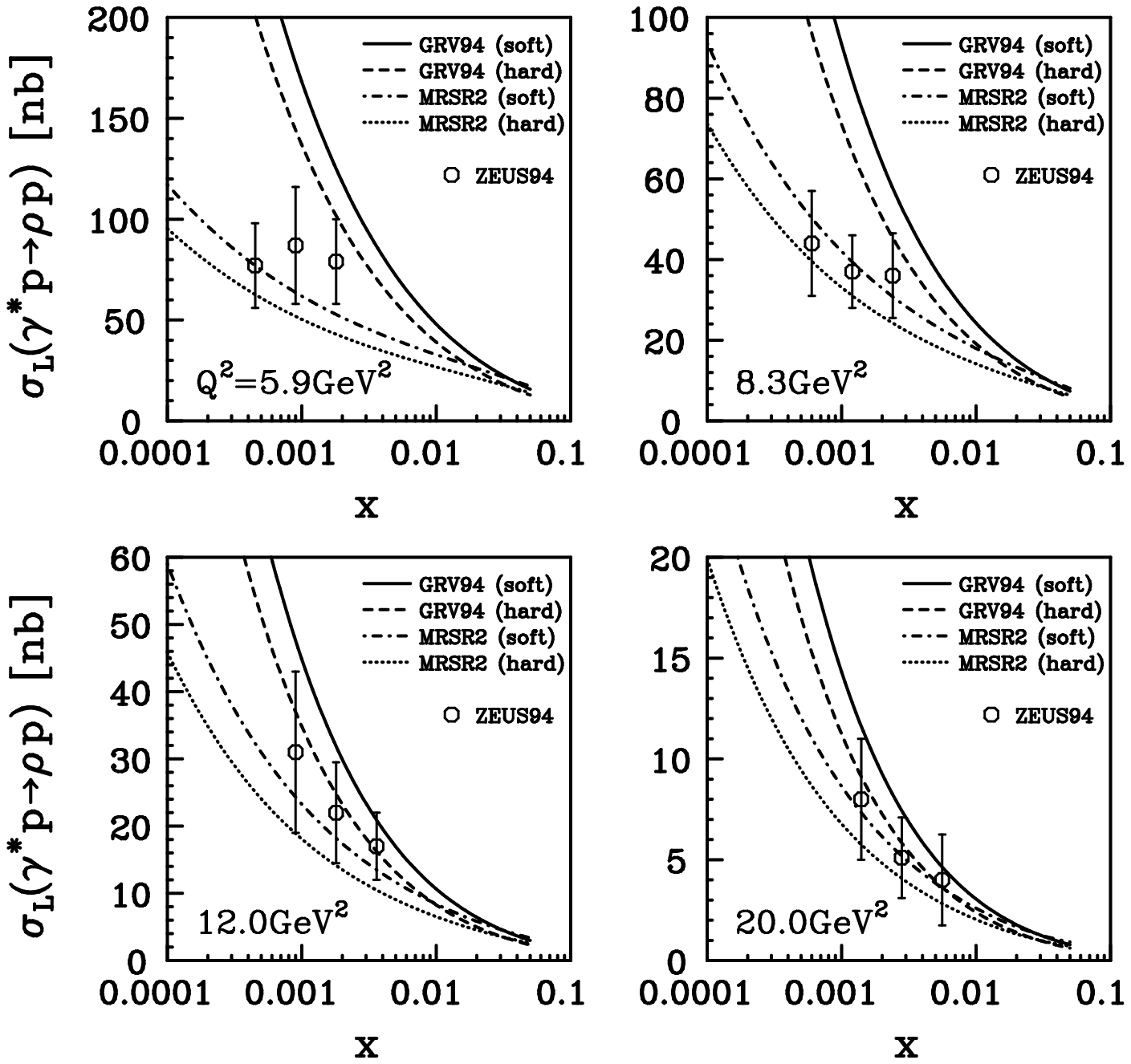}{6.0}
{The longitudinal $\rho^o$ electroproduction cross section,
$\sigma(\gamma_L^*+p \rightarrow \rho^o+p)$ for two extremal parameterizations
of the gluon density \protect\cite{GRV,MRS} and for two different 
$\rho$-meson wave functions in comparison with preliminary ZEUS data 
\protect\cite{ZEUS94}.}

The comparison of our predictions\footnote{A respective FORTRAN program is
available by request from koepf@mps.ohio-state.edu or via the WWW at
http://www.physics.ohio-state.edu/\~{}koepf.} with the most recent 
experimental data is shown in Fig. 14.  
In the kinematic domain were our approach is expected to be applicable, $x 
\lesssim 0.01$ and/or $W \gtrsim 30$ GeV, our predictions agree with the data 
within the spread through the various parameterizations for the gluon density 
and the uncertainty which stems from the vector meson's wave function.  Note, 
in particular, that as $Q^2$ increases the energy dependence of the 
preliminary ZEUS data \cite{ZEUS94} approaches more and more the hard physics 
limit, $\sigma \propto W^{0.7 - 0.8}$, which is very different form the 
soft Pomeron prediction \cite{pom}, $\sigma \propto W^{0.22 - 0.32}$.
This could indicate a transition form soft to hard physics in the $Q^2$ range 
depicted in Fig. 14.

There are two reasons why our predictions should not really reproduce the data 
very well at smaller $Q^2$.  Firstly, smaller $Q^2$ correspond to larger 
transverse distances, and hence the pQCD approach outlined here loses some 
of its validity.  Secondly, at very small $x$, the increase of these 
cross sections with energy is restricted by the unitarity of the $S$-matrix,
and even more stringent restrictions follow from the condition that the 
leading twist term should be significantly larger than the next to leading 
twist term \cite{hepph}.  The kinematical region where this limit 
becomes important moves to larger $x$ for decreasing $Q^2$.
However, whether the softer energy dependence of the cross sections at the
smaller $Q^2$ is really due to the unitarity limit slow-down is unclear
at the moment.  Further work is in progress in that realm \cite{prep}.

The $Q^2$ dependence of the cross section is commonly parameterized
through a quantity $\alpha$, where (for fixed $W$)
\begin{equation}
\sigma(\gamma^*+p \rightarrow \rho^o+p)~\propto~Q^{-2\alpha} \ ,
\ .
\label{eq28}
\end{equation}
The various experiments yield $\alpha = 2.1 \pm 0.4$ \cite{ZEUS},
$\alpha = 2.4 \pm 0.3$ \cite{ZEUS94} and $\alpha = 2.5 \pm 0.5$ \cite{H1}
at $\langle Q^2 \rangle \approx 12$ GeV$^2$ and $\langle W\rangle \approx 
80$ GeV.  Neglecting the Fermi motion corrections and the ``rescaling
of hard processes", our
theoretical predictions yield $\alpha \approx 3.3$ without the
corrections, while we
find $\alpha \approx 2.6$ if we take the quark motion and ``rescaling of hard 
processes" into account.  To evaluate the correction factor $T(Q^2)$ of
Eq. (\ref{eq13}), we again used the wave function $\phi_\rho(z,k_t) 
\propto \exp\!\left(-{A k_t^2\over z(1-z)} \right)$ with an average 
transverse quark momentum of $\langle k^2_t \rangle = 0.18$ GeV$^2$ 
as extracted from a QCD sum rule analysis \cite{halperin}.  Hence,
our predictions agree with the measurements only if the Fermi motion 
corrections and the ``rescaling of hard processes" are taken into account.  
This underlines our claim \cite{hepph} that the $Q^2$ dependence of those 
cross sections could eventually be used to probe the transverse momentum 
distributions within the produced vector mesons.  However, at present, the 
data are still far too crude to extract conclusive information on this
quantity.  Note, furthermore, that our prediction refers to the $Q^2$ 
dependence of the longitudinal cross section, $\sigma_L$, while the 
experimental values listed in the above correspond to the $Q^2$ dependence 
of the total cross sections, $\sigma=\sigma_T + \epsilon \sigma_L$.

\section{Conclusions}
\label{VI}

In this work, we focused the QCD analysis of Refs. \cite{Brod94} and
\cite{hepph} on heavy quarkonium ($J/\psi$ and $\Upsilon$) photo- and 
electroproduction, and we extended the respective formalism, which in 
Refs. \cite{Brod94} and \cite{hepph} was applied to the production of 
longitudinally polarized vector mesons only, to transverse
polarizations as well.

For non-asymptotic momentum transfers, the respective hard amplitude is 
sensitive to the transverse momentum distribution in the $q\bar q$ light-cone 
wave function of the leading Fock component in the produced vector
meson.  This 
leads to a suppression of the asymptotic predictions, i.e., to an interplay 
between the quark(antiquark) momentum distribution in the vector meson and the 
$Q^2$ dependence of the corresponding cross section.  We derived the respective
expressions for the Fermi motion suppression factor, $T(Q^2)$ of Eqs. 
(\ref{eq0g}) and (\ref{eq13}), and the relative enhancement of the transverse 
cross section, $R(Q^2)$ of Eqs. (\ref{eq0f}) and (\ref{eq14}), to
leading order in ${1 \over Q^2+4m^2}$.

The evaluation of these factors required a detailed study of the 
vector meson's 
$q\bar q$ light-cone wave function.  Motivated by the large value of the quark 
mass in heavy quarkonia, we started from conventional non-relativistic 
potential models, which we critically examined and confronted with QCD 
expectations.  In particular for the $J/\psi$ meson, our numerical analysis 
yields a significant value for the high momentum component in the respective 
wave functions, visible in the lower part of Fig. 5, and a significant 
contribution of the ``relativistic region" ${v\over c}\geq 1$ to the integral 
$\int\!d^3k\,\phi_V(k)$, displayed in Fig. 6.  This is in line with large 
relativistic corrections to the corresponding bound state equations 
\cite{buchmueller}.  These large relativistic effects question the feasibility
of a description of heavy meson production in high energy processes based on a 
non-relativistic ansatz. This is a very important result which should have 
consequences far beyond the scope of diffractive vector meson production, and
it indicates that the $J/\psi$ meson is not really a non-relativistic system!
We therefore designed an interpolation for the wave function of heavy 
quarkonia which smoothly matches the results obtained from non-relativistic 
potential models with QCD predictions at short distances.

We then used the latter to evaluate the finite $Q^2$ corrections for 
diffractive $J/\psi$ as well as $\Upsilon$ production.  We find fairly good 
agreement of our predictions with the $J/\psi$ data, and we predict a 
measurable production of $\Upsilon$ mesons at HERA -- especially after a 
luminosity upgrade.  We also update our comparison of longitudinal $\rho^o$
electroproduction with the data, putting special emphasis on preliminary 
ZEUS '94 data \cite{ZEUS94} that became available only recently.

The discussion in this work affirms that hard diffractive vector meson
production is exactly calculable in QCD in the same sense as leading twist
deep inelastic processes.  This holds if only short distances contribute,
which is the case for heavy flavors or production of longitudinally polarized
$\rho^o$ at large $Q^2$.  The respective amplitude is expressed through the
distribution of bare quarks in the vector meson and the gluon distribution
in the target.  This is qualitatively different from an application of
the constituent quark model to these processes, as in Refs. \cite{Ryskin} and
\cite{Ryskin2}.  On the other hand, it makes these processes an
ideal laboratory to study the $q\bar q$ leading Fock state in vector mesons.

\section*{Acknowledgments}

We would like to thank E. Braaten, S.J. Brodsky and R.J. 
Perry for a number of fruitful discussions.  This work was supported in part 
by the Israel-USA Binational Science Foundation under Grant No.~9200126, by 
the U.S. Department of Energy under Contract No. DE-FG02-93ER40771, and by the 
National Science Foundation under Grants Nos. PHY-9511923 and PHY-9258270.
Two of us (L.F. and M.S.) would like to thank the DESY theory division for its
hospitality during the time when part of this work was completed.

\end{document}